\documentclass[aps, preprint, groupedaddress, floatfix]{revtex4}

\usepackage{epsfig}
\usepackage{graphicx}
\usepackage{amsmath}

\newcommand{\sto}{SrTiO$_3$ }
\newcommand{\lao}{LaAlO$_3$ }
\newcommand{\stolao}{SrTiO$_3$/LaAlO$_3$ }

\begin{document}

\title{A First-Principles Study of the Electronic Reconstructions of 
$\textrm{LaAlO}_3/\textrm{SrTiO}_3$ Heterointerfaces and Their 
Variants}

\author{Hanghui~Chen$^{1,3}$ Alexie Kolpak$^{2,3}$ and Sohrab
Ismail-Beigi$^{1,2,3}$}

\affiliation{
 $^1$Department of Physics, Yale University, New Haven,
Connecticut, 06511, USA\\
 $^2$Department of Applied Physics, Yale
University, New Haven, Connecticut, 06511, USA\\
 $^3$Center for Research on Interface Structures
and Phenomena (CRISP), Yale University, New Haven, CT 06511, USA}

\date{\today}

\begin{abstract}

We present a first-principles study of the electronic structures and
properties of ideal (atomically sharp) LaAlO$_3$/
SrTiO$_3$ (001) heterointerfaces and their variants such as a new class of 
quantum well systems. 
We demonstrate the insulating-to-metallic
transition as a function of the LaAlO$_3$ film thickness in these systems.
After the phase transition,
we find that conduction electrons are bound to the $n$-type interface
while holes diffuse away from the $p$-type interface, and we explain this
asymmetry in terms of a large hopping matrix element that is unique to
the $n$-type interface. We build a tight-binding model based on these
hopping matrix elements to illustrate how the conduction electron gas
is bound to the $n$-type interface. Based on the `polar catastrophe'
mechanism, we propose a new class of quantum wells at which we can
manually control the spatial extent of the conduction electron gas.
In addition, we develop a continuous model to unify the
LaAlO$_3$/SrTiO$_3$ interfaces and quantum wells and predict the
thickness dependence of sheet carrier densities of these
systems. Finally, we study the external field effect on both
LaAlO$_3$/SrTiO$_3$ interfaces and quantum well systems. Our 
systematic study of the electronic reconstruction of 
LaAlO$_3$/SrTiO$_3$ interfaces may serve as a guide to 
engineering transition metal oxide heterointerfaces. 

\end{abstract}

\maketitle

\section{Introduction}

With the advance of techniques to control thin film growth on the
atomic scale, the study of epitaxial oxide heterostructures is a
rapidly developing area of materials
science \cite{Dagotto-Science-2007}.  Due to the ability to produce a
well-defined single terminated substrate
surface \cite{Kawasaki-Science-1994}, interfaces that are nearly
atomically sharp can be formed.  In many cases, the properties of
these interfaces turn out to be much richer than those of the bulk
constituents.  One of the most interesting examples is the
(001) interface between LaAlO$_3$ (LAO) and SrTiO$_3$ (STO), both constituents 
of which are conventional band insulators in the bulk. Among the
intriguing phenomena observed at this interface is the presence of a
high-mobility quasi two-dimentional electron gas
\cite{Hwang-Nature-2004}, which is sensitive to the
LaAlO$_3$ film thickness on the nanoscale \cite{Mannhart-Science-2006,
Huijben-NatMat-2006, Bell-APL-2009}, and can be tuned via an external
field \cite{Mannhart-Science-2006, Triscone-Nature-2008}. In addition,
both magnetism \cite{Brinkman-NatMat-2007} and superconductivity
\cite{Mannhart-Science-2007} have been observed at LaAlO$_3$/SrTiO$_3$
interfaces.

Due to these interface properties, the LaAlO$_3$/SrTiO$_3$ system
shows great promise for the development of novel applications in
nano-scale oxide electronics \cite{Cen-NatMat-2008, Cen-Science-2009}.
However, though significant efforts have been made in both theory
\cite{Pickett-PRB-2006, Pickett-PRB-2008, Freeman-PRB-2006,
  Albina-PRB-2007, Tsymbal-PRL-2009, Kelly-EPL-2008, Demkov-PRB-2008}
and experiment \cite{Willmott-PRL-2007, Kala-PRB-2007,
  Herranz-PRL-2007, Siemons-PRL-2007, Sing-PRL-2009,
  Salluzzo-PRL-2009, Thiel-PRL-2009}, the origin of the new phases
emerging at the LaAlO$_3$/SrTiO$_3$ interface has not been completely 
resolved yet.
In part, this is due to the fact that more than one mechanism may play
a role in determining the behavior, and furthermore, the dominant
mechanism can depend on external conditions
\cite{Eckstein-NatMat-2007}.

One mechanism that is generally believed to play a role in the
formation of the two-dimensional electron gas is known as the polar catastophe.
This mechanism is a direct result of the charge discontinuity that
occurs at an abrupt interface between a nonpolar (SrTiO$_3$) and a
polar (LaAlO$_3$) material.  Both materials are perovskite oxides with
an $ABO_3$ structure that forms alternating planes of $AO$ and $BO_2$
stacked along the (001) direction.  Consequently, there are two types
of LaAlO$_3$/SrTiO$_3$ interfaces along this direction: TiO$_2$/LaO
(known as the $n$-type interface) and SrO/AlO$_2$ (the $p$-type
interface) \cite{Hwang-Nature-2004, Hwang-NatMat-2006}.  In the ionic
limit, SrTiO$_3$ is composed of charge neutral atomic layers
(SrO)$^0$ and (TiO$_2$)$^0$, while LaAlO$_3$ consists
of positively charged (LaO)$^{+}$ and negatively charged
(AlO)$^{-}$ atomic layers.  When LaAlO$_3$ is grown epitaxially
on the SrTiO$_3$ substrate along the (001) direction, an internal
electric field directed from (LaO)$^+$ to (AlO$_2$)$^-$ is formed
through LaAlO$_3$, resulting in a potential difference that,
mathematically, diverges with increasing LaAlO$_3$ thickness.  In
reality, to offset the diverging potential, an electronic
reconstruction is expected to occur, with electrons transferring from
the film surface (or a $p$-type interface) to the $n$-type interface.
The polar catastrophe mechanism \cite{Hwang-Nature-2004,
Hwang-NatMat-2006} has been confirmed in density functional theory
(DFT) calculations \cite{Demkov-PRB-2008, Chen-PRB-2009,
Pickett-PRL-2009, Son-PRB-2009}, and could be responsible for the
presence of the two-dimensional electron gas at the atomically sharp 
$n$-type interface (i.e., with no defects).

However, the polar catastrophe mechanism alone can not explain all of
the electronic properties of the LaAlO$_3$/SrTiO$_3$ interface, such
as the observed confinement of the transferred electrons within
several nanometers of the $n$-type
interface \cite{Basletic-NatMat-2008}. In previous
work \cite{Chen-PRB-2009}, we have shown that the electrons are bound
to the $n$-type interface as a result of the chemical environment of
the interface, which produces a large hopping matrix element between
La and Ti.

In this paper, we use both DFT and phenomenological modeling to
provide a more detailed and complete picture of the electronic
reconstruction at ideal LaAlO$_3$/SrTiO$_3$ interfaces.  We predict the
behavior of the sheet carrier density as a function of the LaAlO$_3$
thickness by constructing a continuous model which approximates the
LaAlO$_3$ as a homogeneous media.  We also use both DFT calculations
and model calculations to give a simple yet quantitative picture of
the external field effect that has been experimentally realized at the
$n$-type interface \cite{Mannhart-Science-2006,
Triscone-Nature-2008}. In addition, we propose a new class of
quantum well systems, based on the polar catastrophe mechanism, at
which the spatial extent of the two-dimensional electron gas can be manually
controlled.

The remainder of the paper is organized as follows. In Section 
\ref{method}, we discuss the technical details of our DFT 
calculations. We present
our first-principles results in Section \ref{DFT}. 
Section \ref{DFT-symmetric} gives a
brief discussion on the symmetric superlattices.  In \ref{DFT-polar}, we briefly
describe the polar catastrophe at various types of interfaces, and we
introduce the new quantum well systems.  Section \ref{DFT-bound} discusses the
computation of the on-site and hopping matrix elements and a
tight-binding model. An exceptionally large Ti-La hopping is found at 
the $n$-type interface and its origin is explained.
Section \ref{oxygen-repulsion} studies the behavior 
of oxygen vacancies at the interfaces. Section \ref{DFT-thickness} 
examines the thickness
dependence of the sheet carrier density. Simulations of the external
field effect are presented in \ref{DFT-external}.  
In Section \ref{model}, we present a
continuous model which gives a simple yet quantitative
description of both the sheet carrier density and the external field
effect.  We conclude in Section \ref{conclusion}.

\section{Method Details}
\label{method}

Our calculations are performed using density functional theory within 
the \textit{ab initio} supercell plane-wave approach \cite{Payne-RMP-1992},
with the code PWscf in the Quantum-ESPRESSO package
\footnote{See http://www.quantum-espresso.org}. We employ
the local density approximation (LDA) \cite{Kohn-PR-1965} and
ultrasoft pseudopotentials \cite{Vanderbilt-PRB-1990}, which include
semi-core states in Sr, Ti and La. The reference configurations of
the pseudopotentials are: Sr $4s^24p^65s^2$ ($r^s_c=2.0 a_0,
r^p_c=1.8 a_0$), Ti $3s^23p^63d^14s^2$ ($r^s_c=r^p_c=r^d_c=1.8
a_0$), La $5s^25p^65d^16s^{1.5}6p^{0.5}$ ($r^s_c=r^d_c=2.2 a_0,
r^p_c=2.0 a_0)$, Al $3s^23p^1$ ($r^s_c=1.8 a_0, r^p_c=1.82 a_0$)
and O $2s^22p^4$ ($r^s_c=r^p_c=1.3 a_0$), where $a_0$ is the Bohr
radius. The plane wave basis energy cutoff and charge cutoff are
30 Ry and 180 Ry, respectively. We use a Gaussian smearing width
of 5 mRy when sampling the Brillouin zone. The $k$-grid sampling
of the Brillouin zone is $10\times10\times 1$ where the $z$-axis
is orthogonal to the interface. The convergence of the total
energy and total charge density has been checked with $k$-grids of
up to $20\times20\times1$. 
Periodic copies are separated by $\sim$15~\AA~of vacuum.
We also extend the vacuum to $30$~\AA~to check the 
convergence. The effect of the artifical electric fields 
in the vacuum due to the periodic boundary condition, which
turns out to be quite small, is 
discussed in Appendix \ref{appendix-PBC-intrinsic}.  
The force convergence threshold is 26 meV/\AA. In some key 
results we reduce the threshold to 13 meV/\AA~to check the convergence.

In all our calculations, the interfaces are along the (001) direction
so that the $z$ axis is perpendicular to the interface. The
$x$ and $y$ directions of the simulation cell are subject to periodic
boundary conditions and their lengths are fixed to the theoretical lattice
constant of SrTiO$_3$ $a=3.85$~\AA~(1.5\% smaller than
the experimental value). The atomic coordinates are relaxed 
\footnote{Some atoms are fixed in order to
simulate the bulk-like substrate. This is specified in each
section.} until every force component is smaller than the
convergence threshold. The detailed configurations for different calculations
are specified in each section below.

\section{DFT results}
\label{DFT}

\subsection{Symmetric double $n$-type and $p$-type superlattices}
\label{DFT-symmetric}

\begin{figure}[]
\includegraphics[angle=-90,width=13cm]{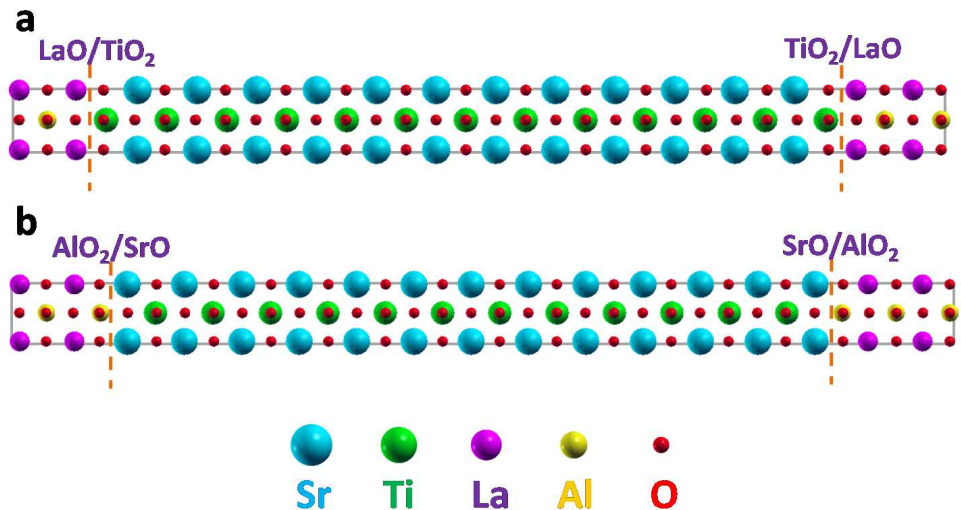}
\caption{\label{fig:symif} Schematics of symmetric superlattices.
\textbf{a)} The double $n$-type superlattice. \textbf{b)} The double
$p$-type superlattice. The interface is highlighted by the dashed
line.}
\end{figure}

The simplest approach to studying the $n$-type or $p$-type
\stolao interface is to use a symmetric
superlattice approach
\cite{Pickett-PRB-2006, Pickett-PRB-2008, Freeman-PRB-2006, Albina-PRB-2007, 
Gemming-Mater-2006, Tsymbal-JAP-2008, Tsymbal-PRL-2009, Satpathy-PRL-2008, 
Kelly-EPL-2008}, as exemplified by Fig.~\ref{fig:symif}, which shows 
a simulation cell
without vacuum in which a number of \sto unit cells are adjacent to a
number of \lao unit cells and periodic boundary conditions are imposed.
Therefore, a non-stoichiometric
supercell contains two identical \stolao interfaces:
{\it e.g.} in Fig.~\ref{fig:symif}a, we have 
an additional TiO$_2$ layer in \sto and an additional LaO layer 
in LaAlO$_3$, and in Fig.~\ref{fig:symif}b, there is an additional 
SrO layer in \sto and an additional AlO$_2$ in LaAlO$_3$.
In this way, the superlattice contains two $n$-type interfaces
(TiO$_2$/LaO) or two $p$-type interfaces (SrO/AlO$_2$).

The advantage of the superlattice approach is that no vacuum is
needed in the simulation cell, making computation easier.
However, due to the imposed
symmetry and non-stoichiometry of the LaAlO$_3$ film, this geometry 
does not result in a polar field, so the evolution of the polar
catastrophe can not be modeled.  In addition, the non-stoichiometry of the
LaAlO$_3$ also imposes a fixed carrier doping: in the ionic limit, an LaO
(AlO$_2$) layer has a charge of +1 (-1), and there is one extra electron
 (hole) present in the conduction (valence) band of LaAlO$_3$, 
which is shared evenly by the
two interfaces.  Therefore each interface is doped by 0.5 electrons
($n$-type) or 0.5 holes ($p$-type) per two-dimensional unit cell. 
These values are precisely those
needed to fully compensate the polar field of LaAlO$_3$
\cite{Hwang-NatMat-2006}, so the symmetric supercell approach is
equivalent to studying the properties of the interfaces when the LaAlO$_3$
film is very thick (infinite thickness limit).

\begin{figure}[]
\includegraphics[angle=-90,width=12.5cm]{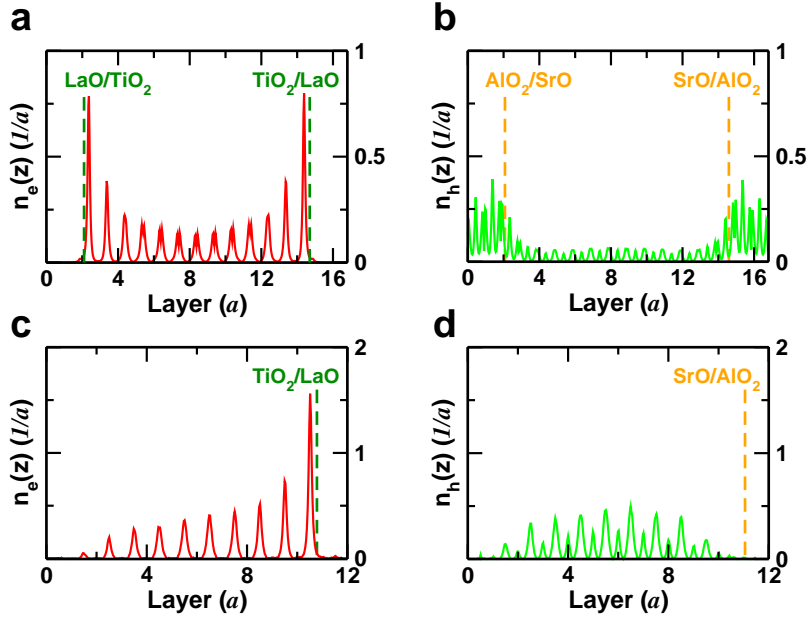}
\caption{\label{fig:chargespatial}
Transferred electron and hole densities
integrated over the $xy$ plane. In each panel, the integral of the
conduction electron/hole density is normalized to unity. The layers are
measured in units of the SrTiO$_3$ lattice constant $a$.
\textbf{a)} The symmetric $n$-type superlattice. The electrons are
bound to the $n$-type interface. \textbf{b)} The symmetric $p$-type
superlattice. The holes reside in both the SrTiO$_3$ and LaAlO$_3$
layers. \textbf{c)} The $n$-type interface with 4 u.c. of LaAlO$_3$. 
The electrons decay away from the $n$-type interface. 
\textbf{d)} The $p$-type interface with 5 u.c. of LaAlO$_3$. The
holes, driven by the polar
field through LaAlO$_3$, diffuse into the \sto substrate. 
Appendix \ref{appendix-density} explains in detail how to calculate these 
transferred charge densities.}
\end{figure}

We briefly present results on the symmetric interfaces
 and highlight some key observations and questions that we
will answer in later sections.  We show in Fig.~\ref{fig:chargespatial}a
the $xy$-plane integrated \textit{conduction}
electron density for the double $n$-type superlattice, 
 and in Fig~\ref{fig:chargespatial}b the hole density for the double
$p$-type superlattice. The method of how to calculate the transferred charge
density is detailed in Appendix \ref{appendix-density}.
The integrated densities are 0.49
electrons per $n$-type interface and 0.49 holes per $p$-type interface, 
showing that the ionic picture for LaAlO$_3$ is highly accurate,
and that one should expect 0.5 electrons (holes) at the fully
compensated $n$-type ($p$-type) interface for very thick LaAlO$_3$ films.

The density profiles in Fig.~\ref{fig:chargespatial} display an 
interesting asymmetry. 
One can characterize the electrons in the double $n$-type system as 
being bound to the interface. As the figure shows, they are confined in 
the \sto, with the density decreasing away from the interface.
On the other hand, in the double $p$-type system, the
holes reside in both the LaAlO$_3$ and SrTiO$_3$, do not show a 
strong preference for the interface itself, and have a slightly higher 
amplitude in the LaAlO$_3$.

Our calculated band
offsets place the LaAlO$_3$ valence band maximum
$\simeq$~0.1 eV above the SrTiO$_3$ valence band maximum ($p$-type) and the
LaAlO$_3$ conduction band minimum $\sim$~2 eV above the SrTiO$_3$ conduction
band minimum ($n$-type) \cite{Albina-PRB-2007, Pickett-PRB-2008}. {\it A
  priori}, one would expect that any introduced carriers would occupy
the most energetically favorable band edge available, {\it i.e.} holes would
migrate to the LaAlO$_3$ valence band edge and electrons to the 
SrTiO$_3$ conduction band edge. Furthermore, 
the minimization of kinetic energy would lead to a
relatively uniform distribution of carriers. In
Fig.~\ref{fig:chargespatial}, we see that, overall, the holes 
prefer to be in the LaAlO$_3$ at the $p$-type interface, 
while the electrons reside in
SrTiO$_3$ at the $n$-type interface. For the $p$-type interface, 
the small value of the band offset
and the large hole density result in some ``leakage'' into the SrTiO$_3$ 
valence band states.

However, the binding of the electrons at the $n$-type interface cannot 
be explained by the above arguments. As no polar fields are present in this
system to create a potential that will bind the electrons, what mechanism
overcomes the kinetic cost inherent in localizing them to the
interface?  Whatever mechanism is present is absent for the
holes, whose spatial profile is relatively uniform in both materials.
We provide the answer to this fundamental question in Section
\ref{DFT-bound}, when we discuss the tight-binding hopping matrix
elements across both interfaces.

Before we move onto the next section, we would like to give a short 
discussion on the valence band offset (VBO) of LaAlO$_3$/SrTiO$_3$ interfaces. 
The VBO of the $n$-type interface has been calculated by several groups 
\cite{Albina-PRB-2007, Lee-MRS-2007, Pickett-PRB-2008, Cen-NatMat-2008, 
Satpathy-PRL-2008}. Although all of the computed values place 
the conduction band 
edge of LaAlO$_3$ significantly higher than that of SrTiO$_3$ 
at the $n$-type interface, the large range of computed VBO values (shown 
in Table I), is puzzling. To investigate this issue, we use 
two different approaches to calculate the VBO of the 
$n$-type interface and the $p$-type interface. 
The valence band offset is defined by:

\begin{equation}
\label{vbo}
E_{\textrm{VBO}}=E_{\textrm{V}}^{\textrm{LAO}}-E_{\textrm{V}}^{\textrm{STO}}
\end{equation} 
where $E_{\textrm{V}}^{\textrm{LAO}}$ and $E_{\textrm{V}}^{\textrm{STO}}$ are 
the valence band edges of LaAlO$_3$ and SrTiO$_3$, respectively. The simulation
cell is a symmetric nonstoichiometric superlattice with 12.5 layers of 
SrTiO$_3$ and 4.5 layers of LaAlO$_3$. The first method 
to determine the VBO is to analyze the local density of states, referred to 
as the LDOS approach \cite{Balereschi-JPD-1998, Albina-PRB-2007}. 
The other method is to use the 
macroscopic average potential, denoted further as the MA approach 
\cite{Baldereschi-PRL-1988, Baroni-PRB-1991, Albina-PRB-2007}. 
We compare our results with other theoretical 
calculations and the recent experiment in Table \ref{tab:vbo}.

\begin{table}[h]
\caption{\label{tab:vbo} The valence band offset (VBO) of LaAlO$_3$/SrTiO$_3$
$n$-type and $p$-type interfaces in eV. Each value is followed 
by the method used to determine it in parentheses.}
\begin{center}
\begin{tabular}{c|c|c}
\hline
\hline 
                    & $n$-type (TiO$_2$/LaO)    & $p$-type (SrO/AlO$_2$) \\
\hline
present study       & -0.47 (MA)  -0.39 (LDOS) & -0.02 (MA)  0.11 (LDOS) \\ 
\hline
Ref.~\cite{Albina-PRB-2007}& 0.51 (MA) 0.39 (LDOS) & 0.19 (MA) 0.10 (LDOS) \\ 
\hline
Ref.~\cite{Cen-NatMat-2008}&  1.1  & NA\\
\hline 
Ref.~\cite{Pickett-PRB-2008}& -0.15 (LDOS) & NA\\
\hline
Ref.~\cite{Satpathy-PRL-2008} & -0.9 (LDOS) & NA\\
\hline
Ref.~\cite{Lee-MRS-2007} & -0.1 (MA) & NA\\
\hline
Expt.~\cite{Segal-PRB-2009} & -0.35$\pm$0.18 & NA\\
\hline
\hline
\end{tabular}
\end{center}
\end{table}

We can see from Table \ref{tab:vbo} that for the $n$-type interface 
not only the magnitude of VBO 
differs, but the sign is not unanimous. One possible origin for the 
range of theoretical values is likely that the lattice constants 
have minute differences 
(due to different pseudopotentials), which causes a variation in strain and 
probably affects the alignment of valence band edges. Another possibility is 
that we, as well as a few other groups \cite{Albina-PRB-2007, Lee-MRS-2007, 
Pickett-PRB-2008, Satpathy-PRL-2008}, use symmetric nonstoichiometric 
superlattices, in which the macroscopic 
average potentials are flat in both LaAlO$_3$ and SrTiO$_3$, to perform 
the calculations. However, if a 
stoichiometric LaAlO$_3$ slab is employed \cite{Cen-NatMat-2008} 
so that a net internal electric 
field results, the determined VBO will likely be different from the 
symmetric nonstoichiometric case.
Although our results for the $n$-type interface agree well with the 
available experiments, the value as well as variations of the VBO for 
the $n$-type interface is not a closed subject and needs further work. In 
contrast, the computed values of the VBO for the $p$-type interface are 
in general agreement and $\simeq$ 0.1 eV. 
     
\subsection{The polar catastrophe}
\label{DFT-polar}

\subsubsection{$n$-type, $p$-type and $np$-type interfaces}

\begin{figure}[]
\includegraphics[angle=0,width=13cm]{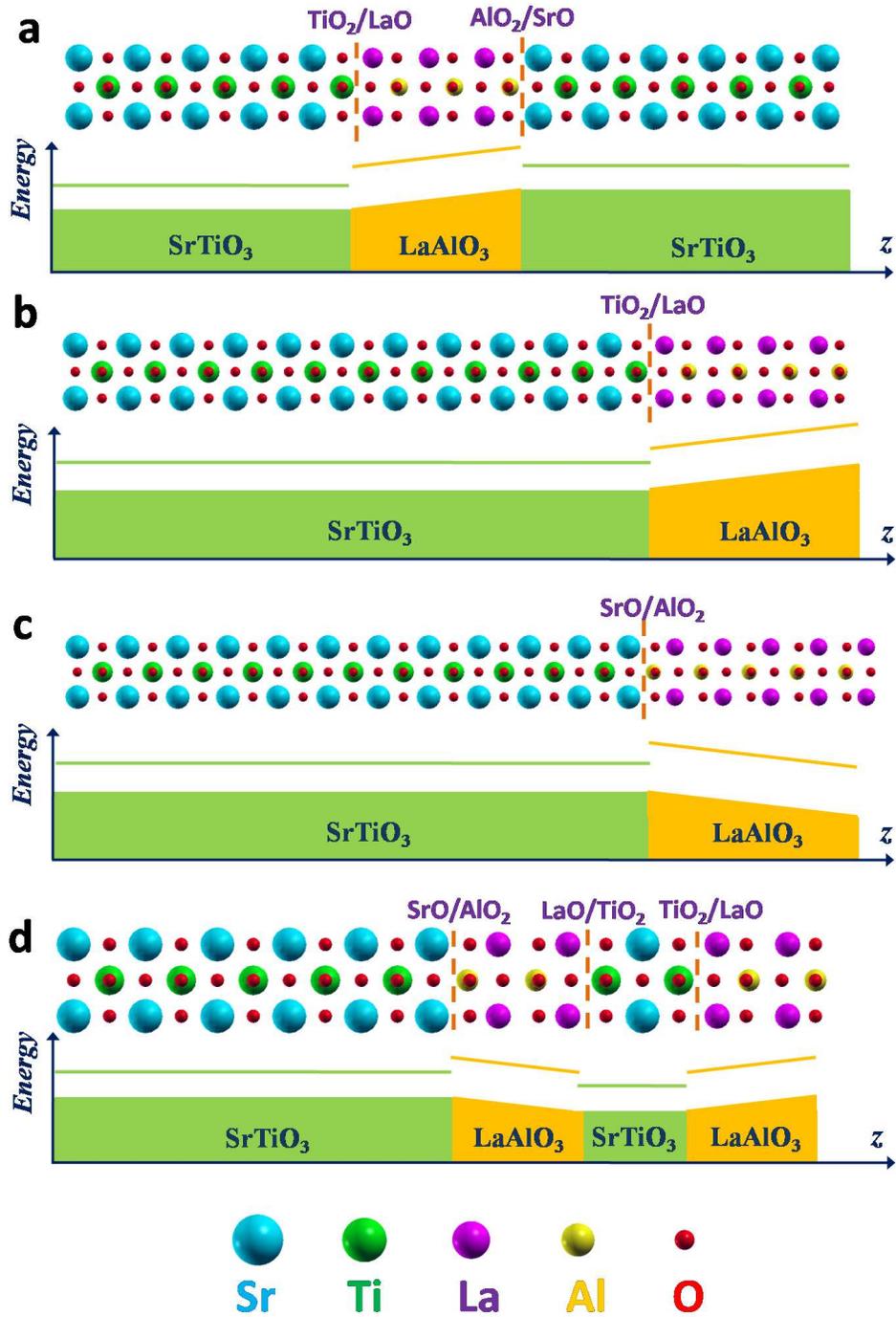}
\caption{\label{fig:interface} Schematics of {the supercells} and energy
diagrams for different types of interfaces. \textbf{a)} The $np$-type
interface. \textbf{b)} The $n$-type interface. \textbf{c)} The $p$-type
interface. \textbf{d)} The new quantum well systems.}
\end{figure}

As discussed in the Introduction, there are two types of
LaAlO$_3$/SrTiO$_3$ interfaces: the $n$-type (TiO$_2$/LaO) interface and
the $p$-type (SrO/AlO$_2$) interface. If a stoichiometric LaAlO$_3$ layer 
on an SrTiO$_3$ substrate also has a
SrTiO$_3$ capping layer, then we have both $n$-type and $p$-type
interfaces in the same system. Below, we refer to this type of  
geometry as the $np$-type interface.
Experimentally, the $n$-type, $p$-type and $np$-type interfaces have all been
fabricated. In this study, we simulate all  three types of interfaces,
the configurations of which are illustrated in Fig. \ref{fig:interface}.
For the $np$-type interface, we use 5 unit cells of
SrTiO$_3$ to simulate the substrate and another 5 unit cells of
SrTiO$_3$ as a capping layer. For both the $n$-type and $p$-type interfaces,
we use 11 unit cells of SrTiO$_3$ to simulate the substrate. In all cases,
the thickness of LaAlO$_3$ is varied from 1 to 7 unit cells. 
In addition, the first two SrTiO$_3$ unit cells in the substrate are 
fixed at the ideal perovskite positions to simulate the bulk-like interior 
of the substrate \cite{Chen-PRB-2009}.
The termination of the SrTiO$_3$ substrate and capping layers is always 
SrO. As illustrated in our 
previous work \footnote{See the EPACS Document No.
E-PRBMDO-79-R12912 for a detailed discussion of surface effects of
TiO$_2$ termination.}, the TiO$_2$ termination has a surface O-$p$ 
state which is $\simeq$ 0.5 eV higher than the valence band edge of 
bulk SrTiO$_3$, while the SrO termination does not have this surface state. 
Since we are interested in the evolution of 
polar fields of LaAlO$_3$ on very thick SrTiO$_3$ substrates, 
this pure surface state should be avoided when 
simulating the SrTiO$_3$ substrate and thick SrTiO$_3$ capping layers. 
However, for thin SrTiO$_3$ capping layers, the presence of this 
O-$p$ state on the TiO$_2$-terminated surface will lower the critical 
separation, which has recently been discussed \cite{Pentcheva-2009}.

We observe that an insulating-to-metallic transition occurs in our simulations 
when the \lao film reaches a critical thickness, the value of which depends on 
the system geometry.
The phase transition can be explained in terms of an
energy diagram shown in Fig. \ref{fig:interface}. For the
$np$-type and $n$-type interfaces, due to the polar structure of
LaAlO$_3$, the electric field through the LaAlO$_3$ film lifts up the
valence band edge of the LaAlO$_3$ and reduces the energy gap. The
energy gap of the $np$-type interface is the energy difference between
the Ti-$d$ states of the SrTiO$_3$ substrate and the O-$p$ states of the
SrTiO$_3$ capping layer, while the energy gap of the $n$-type
interface is given by the energy difference between the Ti-$d$
states of the SrTiO$_3$ substrate and the O-$p$ states on the surface. For
the $p$-type interface, since the polarity of LaAlO$_3$ is
reversed, the electric field through the LaAlO$_3$ film decreases the
valence band edge of the LaAlO$_3$. However, the energy gap of
the $p$-type interface is not the most relevant quantity to monitor for the
insulating-to-metallic transition. Rather, the LaAlO$_3$ film reduces
the energy difference between the La states on the surface
\footnote{Detailed calculations show that the surface states have
character of La $5d$, $6s$ and $6p$.} and the O $p$-states of the
SrTiO$_3$ substrate, which from now on we call the `La-O energy
difference'. In all three cases, the energy gap ($np$-type and
$n$-type) or La-O energy difference ($p$-type)
diminishes with the LaAlO$_3$ film thickness, finally
disappearing when the insulating-to-metallic transition occurs. The
minimum number of LaAlO$_3$ unit cells necessary to induce this phase
transition is referred to as the `critical separation'.

Our DFT calculations, as well as previous studies 
\cite{Demkov-PRB-2008, Chen-PRB-2009, Pickett-PRL-2009, Son-PRB-2009}, 
support the above schematics. 
The values of the calculated energy gap versus the LaAlO$_3$ thickness
are listed in Table \ref{tab:gap} for the $np$-type, $n$-type and 
$p$-type interfaces. The corresponding `critical separation' is the
smallest thickness of LaAlO$_3$ that makes the interface conducting.
The increasing thickness of the LaAlO$_3$ film reduces the energy gap until it
disappears and the interface becomes metallic. However, the
critical separation depends on the band gap of SrTiO$_3$, which is
underestimated in DFT calculations (the calculated band gap is 1.85
eV and the experimental band gap is 3.2 eV). Taking into account
the difference between the calculated and experimental band gaps,
the realistic critical separation is two more unit cells of LaAlO$_3$ 
in addition to the calculated one, resulting in 5, 6 and 8 u.c. for  
the $np$-type, $n$-type and $p$-type interfaces, respectively.

\begin{figure}[]
\includegraphics[angle=0,width=10cm]{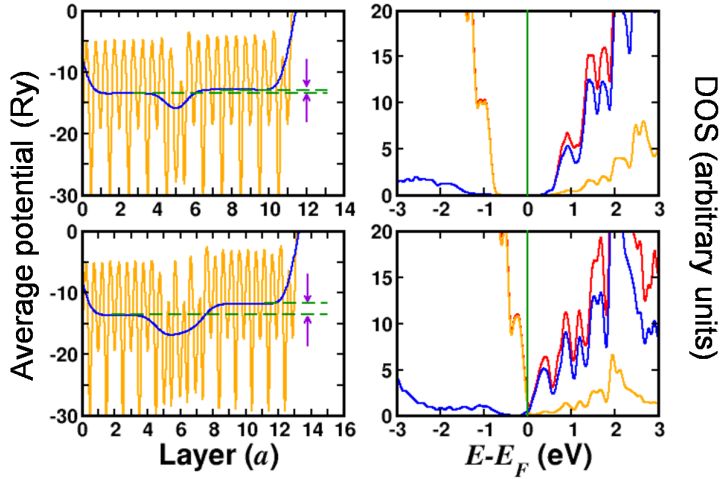}
\caption{\label{fig:potdos}The average potential (left) and density of
states (right) of the $np$-type interface with different thicknesses. 
The upper two panels correspond to the $np$-type interface
with 1 unit cell of LaAlO$_3$, and the lower two panels correspond to
the $np$-type interface with 3 unit cells of LaAlO$_3$. In the right 
panels, the red, blue, 
and orange lines are the total density of states (DOS), the atomic 
projected DOS of Ti-$d$ states,
and the atomic projected DOS of O $p$-states, respectively. The 
vertical green line in the right panels is the Fermi level.}
\end{figure}

In order to give a clearer illustration of how the polar field reduces
the energy gap, we show in the left column of Fig.  \ref{fig:potdos},
the $xy$ planar average electrostatic potential
\cite{Baldereschi-PRL-1988} and the associated macro-averaged smoothed
potential along the direction perpendicular to the interface for the
$np$-type geometry.  The $xy$ planar average potential is obtained by
averaging the raw three-dimensional total potential over the interface
plane. The associated macro-averaged smoothed potential is to convolute
the $xy$ planar average potential with a Gaussian function using the
filting width $\simeq 0.6$\AA. As the figure shows, the macro-averaged smoothed
potential in SrTiO$_3$ is flat, indicating that there is no internal
electric field through the SrTiO$_3$ ({\it i.e.}, it is nonpolar). In
the LaAlO$_3$ layers, the macro-averaged smoothed potential is increasing and
lifts up the valence band edge of the capping SrTiO$_3$.  The energy
difference between the O-$p$ states of the SrTiO$_3$ capping layer and
those of the SrTiO$_3$ substrate increases with increasing LaAlO$_3$
film thickness. The right column of Fig.  \ref{fig:potdos} shows the
corresponding density of states (DOS).  As the upper panel shows,
since the potential increase by 1 unit cell of LaAlO$_3$ is not large
enough to overcome the band gap of SrTiO$_3$, the system is still
insulating. In the lower panel, the potential increase by 3 unit cells
of LaAlO$_3$ is larger than the band gap of SrTiO$_3$, so that the
Ti-$d$ states in the SrTiO$_3$ substrate and O-$p$ states in the
capping SrTiO$_3$ layer overlap. Then the system becomes metallic.

\begin{figure}[]
\includegraphics[angle=-90,width=14cm]{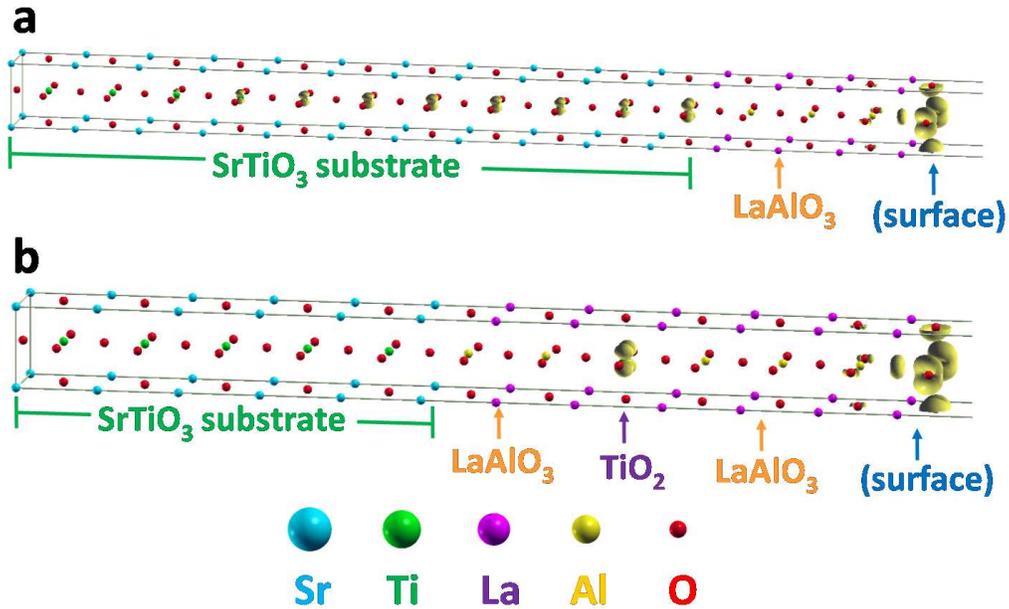}
\caption{\label{fig:ldosf}  3D isovalue surfaces (yellow contour) showing 
the local density of states at the Fermi level for {\bf a)} the 
$n$-type interface and {\bf b)} the new quantum well systems. 
The conduction 
electrons (occupying Ti-$d$ orbitals) extend into the SrTiO$_3$ substrate at 
the $n$-type interface, but are localized in the embedded TiO$_2$ layer 
at the QW. Holes occupy the O-$p$ states on the surface in both cases.}
\end{figure}

Further evidence of the insulating-to-metallic transition is
shown by the local density of states (LDOS) at the Fermi level of the
$n$-type interface in Fig.  \ref{fig:ldosf}a.  From the character
of these states, we can see that we have Ti-$d$ states in the
SrTiO$_3$ substrate and O-$p$ states on the surface.
Fig. \ref{fig:chargespatial} shows the spatial
distributions of the conducting electrons and the holes
integrated over $xy$-plane at the $n$-type and $p$-type interfaces,
respectively. The details of how to extract out these transferred 
charge densities are provided in Appendix \ref{appendix-density}. 
From Fig.  \ref{fig:chargespatial}, it is clear
that the conducting electrons and holes behave very differently
in the SrTiO$_3$ substrate. The conducting electrons, which
occupy Ti-$d$ states, decay away from the $n$-type interface over a
length scale of $\simeq$ 10 unit cells, indicating that the
conducting electron gas is several nm thick. This result
is in qualitative agreement with experimental measurements of the
superconducting electron gas \cite{Mannhart-Science-2007} and other
theoretical calculations \cite{Tsymbal-PRL-2009, Son-PRB-2009}. Unlike the
conducting electrons, the spatial distribution of the holes resembles
that of a particle in a box, implying that the holes are very weakly bound to
the interface. As elucidated in our previous work
\cite{Chen-PRB-2009}, the reason for this asymmetry is that the
conducting electron gas is trapped at the $n$-type interface due
to a large interfacial hopping matrix element. We will discuss
this phenomenon further in Section \ref{DFT-bound}.  

\subsubsection{The quantum well system}
\label{sec:quantumwell}

Though the conduction electrons are bound to the $n$-type interface,
they still spread over several nanometers. The observed
thickness of the electron gas at the $n$-type interface varies widely
in experiments, from nanometers to microns, depending on the growth
conditions \cite{Chen-Adv-2010}. 
Thus it would be desirable to be able to manually control
the thickness of the electron gas.
Based on the polar catastrophe mechanism, we propose a new class
of quantum well systems that will allow this functionality. The
quantum well system, which is illustrated in
Fig. \ref{fig:interface}d, has the following composition: \sto
substrate/2 u.c. \lao/$m$ u.c. \sto/\lao capping layer, where $m$ can
be varied to control the thickness of the conducting electron gas, 
as we discuss below.
Along the (001) direction, this configuration results in three
interfaces: one $p$-type interface between the SrTiO$_3$ substrate and 
the 2 u.c. \lao buffer layer,
and two $n$-type interfaces on both sides of the inserted thin \sto film.  
Both of the
LaAlO$_3$ thin films are stoichiometric in this sytem, while the
embedded SrTiO$_3$ thin film is not. As we have two $n$-type
interfaces, the quantum well systems have one extra TiO$_2$
layer in the inserted SrTiO$_3$ film. 
Like the $n$-type and $np$-type interfaces, the two SrTiO$_3$ unit
cells of the substrate facing the vacuum are fixed to 
simulate a bulk-like substrate
and all the other atoms are fully relaxed.

The reason that this configuration forms a quantum well can also be
explained in terms of an energy diagram.  Fig. \ref{fig:interface}d
shows that the polar fields in the two LaAlO$_3$ thin films 
point in opposite directions. Therefore, the energy of the Ti-$d$
states in the embedded SrTiO$_3$ thin film is lowered relative to
the conduction band edge of the SrTiO$_3$ substrate. (We choose two
unit cells of LaAlO$_3$ as the buffer layer in order to bring 
down the Ti-$d$ states
of the embedded SrTiO$_3$ film into the band gap of the SrTiO$_3$
substrate). As the capping LaAlO$_3$ layer thickens, the energy of
the O-$p$ states eventually becomes higher than that of the Ti-$d$
states of the embedded SrTiO$_3$ film, and hence an
insulating-to-metallic transition occurs. The difference between the
quantum well systems and the $n$-type interface is that once the
insulating-to-metallic transition happens, the conduction electrons
get trapped in the Ti-$d$ states of the embedded SrTiO$_3$ film
rather than those of the SrTiO$_3$ substrate. Therefore, by changing
the thickness of the embedded SrTiO$_3$ film, we can manually
control the spatial extent of the electron gas while holding the 
number of transferred electrons constant. Consequently, we
have two parameters to vary: 1) the thickness of embedded SrTiO$_3$
film to control the spatial extent of electron gas, and 2) the
thickness of LaAlO$_3$ capping layer to control the
insulating-to-metallic transition.  For simplicity, in the
following, we focus on one particular subclass in which the embeded
SrTiO$_3$ is narrowed down to a single TiO$_2$ layer. The thickness of
the LaAlO$_3$ capping layer will be varied. The quantum wells
(QW) in the following refer to this particular subclass.

Like the $p$-type interface, at this new class of quantum well systems, 
the energy gap is
not the most direct quantity to monitor for the insulating-to-metallic
transition. As the LaAlO$_3$ capping layer is thin, the
energy gap of the QW is the energy difference between the Ti-$d$
states of the embedded TiO$_2$ layer and the O-$p$ states of the
SrTiO$_3$ substrate, which remains at a constant energy. The more relevant
quantity is the energy difference between the Ti-$d$ states of the
embedded TiO$_2$ layer and the O-$p$ states on the surface.  We
refer to this quantity as the `Ti-O energy difference' in the
following.  Increasing the thickness of the \lao capping layer
reduces the Ti-O energy difference until it disappears and electron
transfer occurs. The minimum thickness of the LaAlO$_3$ capping layer
that enables this electron transfer is defined as the critical
separation of the QW systems.

Table \ref{tab:gap} shows the Ti-O energy difference and the critical
separation of the QW systems. The critical separation of the QW
coincides with that of the $n$-type interface, reflecting the
fact that the band gap of SrTiO$_3$ is mainly determined by the Ti-$d$
states and the O-$p$ states. Therefore, the absence of Sr atoms does
not change the critical separation. The local density of states
at the Fermi level of the QW are shown in Fig. \ref{fig:ldosf}b.
Unlike the $n$-type interface, at which the filled Ti $d$-states
extend over $\sim$10 unit cells in the \sto substrate, the only
metallic states in the QW systems are localized in the single embedded
TiO$_2$ layer, as anticipated.

\begin{figure}[t!]
\includegraphics[angle=0,width=16cm]{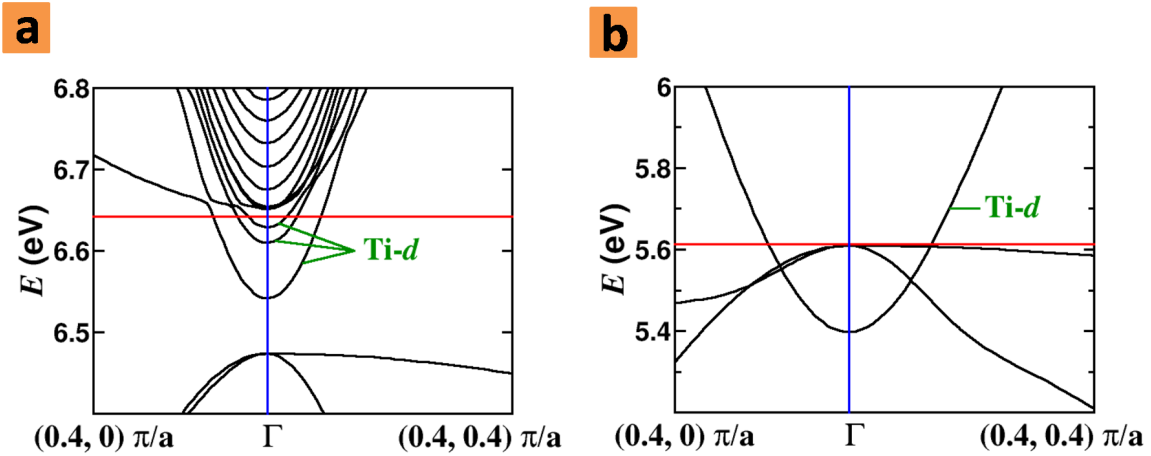}
\caption{\label{fig:bands_compare} A comparison of the band structures 
of {\bf a)} the $n$-type interface and {\bf b)} the new QW.  
The $n$-type interface has 5 unit cells of LaAlO$_3$ and 
the QW has a capping layer of 
6 unit cells of LaAlO$_3$. The red solid line is the Fermi level.}
\end{figure}

This new class of quantum wells displays a number of appealing
properties that are absent at the LaAlO$_3$/SrTiO$_3$ $n$-type
interface. First, the thickness of conduction electrons is controlled
by the inserted SrTiO$_3$ film and can be in principle reduced to
a single atomic TiO$_2$ layer. Second, the electronic
properties of the new quantum wells largely depend on the polar
structure of LaAlO$_3$. Therefore, the SrTiO$_3$ substrate can be
replaced by other materials on which LaAlO$_3$ can be epitaxially
grown, \textit{e.g.} LSAT (La$_{0.29}$Sr$_{0.71}$Al$_{0.65}$Ta$_{0.35}$O$_3$). 
This provides more choices of substrates
on which to grow this new quantum well structure. Finally,
the new quantum well serves as a practical way to test the recently
proposed hypothesis that some electrons do not contribute to transport
due to Anderson localization. Popovi\'{c}
\textit{et al.}~\cite{Satpathy-PRL-2008} argued that since all the two
dimensional states are Anderson localized by disorder, the electrons
that occupy the lowest Ti-$d$ bands do not conduct and therefore the
observed sheet carrier density should be much smaller than 
the 0.5$e$ per two-dimensional unit cell predicted by the polar catastrophe mechanism
\cite{Hwang-NatMat-2006}. Nevertheless, it is not trivial to
distinguish from which band conduction electrons originate. The
thinnest of this new type of quantum wells 
has only one Ti-$d$ band at the Fermi level, 
corresponding to the single embedded TiO$_2$ layer, 
which has a strong two-dimensional
Ti-$d_{xy}$ character. Therefore it should show strong localization and 
not contribute to conductivity. By thickening the embedded SrTiO$_3$, 
one should 
be changing only the extent of the wave functions and see the localization 
properties. Fig. \ref{fig:bands_compare} shows a
comparison of the band structures of the $n$-type interface and
the QW.
These two systems both have approximately 0.1$e$ per two-dimensional unit cell
(from Fig. \ref{fig:charge}). However, the Fermi level of the $n$-type
interface crosses three Ti-$d$ bands, while the QW system has
only one Ti-$d$ band at the Fermi level, as expected.

\begin{table}[h]
\caption{\label{tab:gap} DFT-LDA computed energy gaps and
potential changes versus the number of LaAlO$_3$ layers $i$ for
the $np$-type, $n$-type and $p$-type interfaces ($i$=0 refers to a
pure SrTiO$_3$ substrate). $\Delta$ is the energy gap of the
interface systems ($np$-type and $n$-type) with the corresponding
number of LaAlO$_3$ layers. $\delta$ is the `La-O energy
difference' ($p$-type interface) or `Ti-O energy difference'
(quantum well) with the corresponding number of LaAlO$_3$ layers.
$V_i$ is the macroscopic potential change due to adding the $i$-th
LaAlO$_3$ layer. Note that `La-O energy difference' is not well
defined when $i=0$ (no LaAlO$_3$ layer).}
\begin{center}
\begin{tabular}{ccc|ccc|ccc|ccc}
\hline
\hline & $np$-type& & &$n$-type& & & $p$-type& & & QW&   \\
\hline $i$ & $\Delta$(eV) & $V_i$(eV) &  $i$  & $\Delta$(eV) &
$V_i$(eV) & $i$ & $\delta$(eV) & $V_i$(eV) & $i$ &$\delta$(eV) & $V_i$(eV)\\
\hline 0 & 1.85     &      & 0 & 1.85     &      & 0   & N/A     &        & 0  & 1.85     &     \\
\hline 1 & 1.25     & 0.60 & 1 & 1.71     & 0.14 & 1   & 2.59    &        & 1  & 1.83     & 0.02\\
\hline 2 & 0.55     & 0.70 & 2 & 1.28     & 0.43 & 2   & 1.89    & 0.70   & 2  & 1.30     & 0.53\\
\hline 3 & metallic &      & 3 & 0.57     & 0.71 & 3   & 1.18    & 0.71   & 3  & 0.59     & 0.71\\
\hline   &          &      & 4 & metallic &      & 4   & 0.49    & 0.69   & 4  & metallic &     \\
\hline   &          &      &   &          &      & 5   & metallic&        &    &          &     \\
\hline \hline
\end{tabular}
\end{center}
\end{table}

\subsection{Bound versus unbound carriers at the interfaces}
\label{DFT-bound}

In this section, we provide a microscopic picture of the quantum
states inhabited by the carriers at the $n$-type and $p$-type
interfaces, based on a first-principles extraction of tight-binding
parameters. As reported in our previous work \cite{Chen-PRB-2009},
the main result is that a significant Ti-La interfacial hopping
element unique to the $n$-type interface causes electron to bind
there. No such mechanism is operative at the $p$-type interface,
explaining the delocalization of the holes into the SrTiO$_3$
substrate.  As shown below, the Ti-La hopping is 
significantly enhanced at the $n$-type 
interface due to the relatively large size of La $d$-orbitals as 
well as the spatial proximity of the La and Ti atoms in the 
neighboring atomic planes. Here we provide an expanded and more detailed
explanation and analysis of the hopping elements. 

To calculate the on-site and hopping matrix elements, we employ 
L\"{o}wdin atomic orbitals~\cite{Lowdin-JChem-1950} 
where $\langle \textbf{r}|\alpha \textbf{R}\rangle=
\phi_{\alpha}(\textbf{r}-\textbf{R})$ is a L\"{o}wdin orbital of type
$\alpha$ localized around lattice position $\textbf{R}$.
 However, as
the Kohn-Sham wave functions are Bloch states indexed by wave vectors
\textbf{k}, it is more fruitful to employ Bloch-like superpositions
$\varphi^{\textbf{k}}_{\alpha}(\textbf{r})$ defined as
\begin{equation}
\label{hopping-1} \varphi^{\textbf{k}}_{\alpha}(\textbf{r})
=\frac{1}{\sqrt{N_{\textbf{k}}}}
\sum_{\textbf{R}}e^{i\textbf{k}\cdot\textbf{R}}
\phi_{\alpha}(\textbf{r}-\textbf{R})\,.
\end{equation}
$N_{\textbf{k}}$ is the total number of $k$-points in the Brillouin
zone sampling. To extract a tight-binding model for the Bloch states
at \textbf{k}, we need to calculate the matrix elements $\langle
\varphi^{\textbf{k}}_{\alpha}|H| \varphi^{\textbf{k}}_{\beta}\rangle$.
Making use of the completeness of the Hamiltionian, we obtain:
\begin{equation}
\label{hopping-4} \langle \varphi^{\textbf{k}}_{\alpha}|H| \varphi^{\textbf{k}}_{\beta}\rangle
=\sum_{n}\langle \varphi^{\textbf{k}}_{\alpha}|n\textbf{k}\rangle E_{n\textbf{k}}
\langle n\textbf{k}| \varphi^{\textbf{k}}_{\beta}\rangle
\end{equation}
where $\langle \textbf{r}|n\textbf{k}\rangle=\psi_{n\textbf{k}}(\textbf{r})$
are the actual Bloch eigenstates of the Hamiltonian $H$. The overlap
$\langle\varphi^{\textbf{k}}_{\alpha}|n\textbf{k}\rangle$ can be
recast easily as follows:
\begin{equation}
\label{hopping-5}\langle\varphi^{\textbf{k}}_{\alpha}|n\textbf{k}\rangle
=\sqrt{N_{\textbf{k}}}\int
d\textbf{r}\phi^{*}_{\alpha}(\textbf{r})\psi_{n\textbf{k}}(\textbf{r})
=\langle \alpha \textbf{0}|n\textbf{k}\rangle
\end{equation}
Our final operational formula is
\begin{equation}
\label{hopping-6} \langle \varphi^{\textbf{k}}_{\alpha}|H|
\varphi^{\textbf{k}}_{\beta}\rangle =\sum_{n}\langle
\alpha\textbf{0}|n\textbf{k}\rangle E_{n\textbf{k}} \langle
n\textbf{k}| \beta \textbf{0}\rangle
\end{equation}
where the overlaps $\langle \alpha \textbf{0}| n \textbf{k} \rangle$
are automatically computed and reported by the PWscf code.  
Choosing $\alpha=\beta$ gives the on-site energies while 
$\alpha\ne\beta$ are the hopping elements. The orthonormality 
relation can also be obtained by replacing $H$ with the identity $I$ 
operator:

\begin{equation}
\label{hopping-7} \langle \varphi^{\textbf{k}}_{\alpha}|
\varphi^{\textbf{k}}_{\beta}\rangle =\sum_{n}\langle
\alpha\textbf{0}|n\textbf{k}\rangle \langle
n\textbf{k}| \beta \textbf{0}\rangle=\delta_{\alpha\beta}
\end{equation}
Eq.(\ref{hopping-7}) is a good criterion to check the truncation in 
the infinite summation over the band index $n$. In our calculations, 
we include bands with energies up to 29 eV above the Fermi level, so that  
$|\langle \varphi^{\textbf{k}}_{\alpha}| \varphi^{\textbf{k}}_{\alpha}\rangle|
>0.99$ for all L\"{o}wdin orbitals considered and 
$|\langle \varphi^{\textbf{k}}_{\alpha}|\varphi^{\textbf{k}}_{\beta}\rangle|
<5\times 10^{-4}$ for all the pairs of two different atomic orbitals.

At the $n$-type interface, direct projection of the Bloch states
$\psi_{n\textbf{k}}$ onto the atomic orbitals shows that the
character of the lowest band accommodating the conduction electrons
is mainly Ti-$d_{xy}$ with a small component of
La-$d_{xy}$. The minimum of these occupied bands is at $\Gamma$
$({\bf k}={\bf 0})$. Therefore we build our tight-binding model on
the subspace composed of Ti-$d_{xy}$ and La-$d_{xy}$ orbitals and
calculate the following on-site and hopping matrix elements:

\begin{equation}
\label{hopping-8} H_{00}=\langle
\varphi^{\Gamma}_{\textrm{La-}d_{xy}}|H|
\varphi^{\Gamma}_{\textrm{La-}d_{xy}}\rangle
\end{equation}
\begin{equation}
\label{hopping-9} H_{jj}=\langle
\varphi^{\Gamma}_{\textrm{Ti}^j\textrm{-}d_{xy}}|H|
\varphi^{\Gamma}_{\textrm{Ti}^j\textrm{-}d_{xy}}\rangle
\end{equation}
\begin{equation}
\label{hopping-10} H_{0j}=H^{*}_{j0}=\langle
\varphi^{\Gamma}_{\textrm{La}\textrm{-}d_{xy}}|H|
\varphi^{\Gamma}_{\textrm{Ti}^j\textrm{-}d_{xy}}\rangle
\end{equation}
\begin{equation}
\label{hopping-11} H_{ij}=H^{*}_{ji}=\langle
\varphi^{\Gamma}_{\textrm{Ti}^i\textrm{-}d_{xy}}|H|
\varphi^{\Gamma}_{\textrm{Ti}^j\textrm{-}d_{xy}}\rangle\\
\end{equation}
where `La' in Eq.~(\ref{hopping-8}) and Eq.~(\ref{hopping-10}) is the La
atom in the LaAlO$_3$ layer at the $n$-type interface. $i$ (or $j$) labels the
Ti atoms in the SrTiO$_3$ where $i=1$ (or $j=1$) is in the very TiO$_2$ 
layer at the $n$-type interface, and increasing $i$ (or $j$) refers to the 
TiO$_2$ layers that are further away from the interface.

\begin{figure}[]
\includegraphics[angle=270,width=15cm]{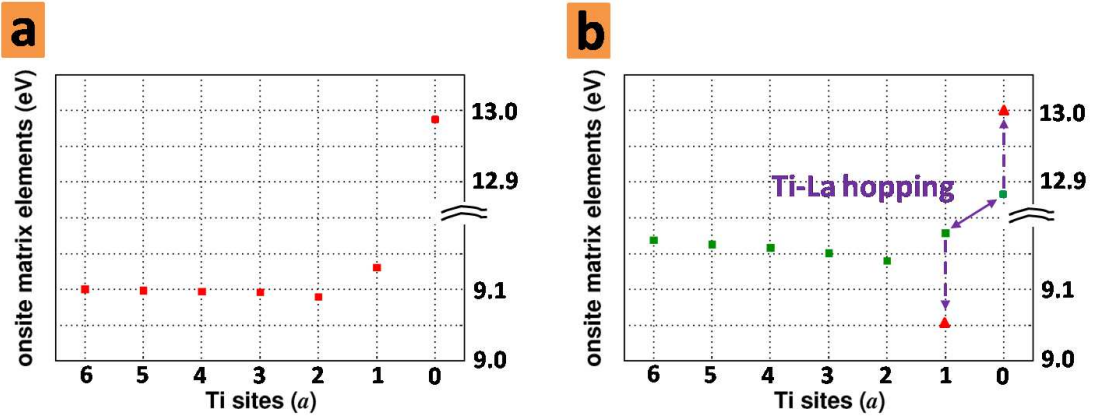}
\caption{\label{fig:onsite} The on-site matrix elements of the $n$-type
interface. The squares are the on-site matrix elements. Site 0 is 
the La atom. Sites 1-6 are the Ti atoms. The triangles are the `new' 
onsite matrix elements after taking into account the hopping effects.
The data can be read from Table \ref{tab:onsite}. $a$ is the lattice 
constant of SrTiO$_3$. {\bf a)} The $n$-type interface with 2 unit cells of 
LaAlO$_3$. {\bf b)} The $n$-type interface with 4 unit cells of LaAlO$_3$. 
The purple arrows illustrate the effect of Ti-La hopping as 
per Eq. (\ref{hopping-14}) and Eq. (\ref{hopping-15}).}
\end{figure}

The $n$-type interface with 2 u.c. of LaAlO$_3$ is insulating (before
the polar catastrophe) and that with 4 u.c. of LaAlO$_3$ is conducting
(after the polar catastrophe). Table~\ref{tab:onsite} shows the
on-site and hopping matrix elements of the $n$-type interface before
and after the polar catastrophe, respectively. The on-site matrix
elements are also shown in Fig.~\ref{fig:onsite}.  We can see that the
polar catastrophe does not result in significant difference in either
the on-site or the hopping matrix elements. Therefore, in the
following we mainly focus on the tight-binding Hamiltonian after the polar
catastrophe, which is more relevant to describe the conduction
electron gas.

Fig.~\ref{fig:onsite} and Table~\ref{tab:onsite} show that 
the on-site matrix element of the first 
Ti atom (closest to the interface) is $not$ the lowest one before 
or after the polar catastrophe; 
instead, the second Ti atom has
the lowest potential. Therefore, one would expect to find the highest
electron density on the second Ti atom, in direct contrast to what is
actually found in Fig.~\ref{fig:chargespatial}c.  However, there is a very
large hopping matrix element between the La and Ti atoms at the
interface, at least 100 times larger than all the other hopping 
matrix elements between Ti atoms. This is due to two factors: one is that 
La-$5d$ orbitals are more extended than Ti-$3d$ orbitals, which leads 
to a larger overlap, and the other is 
that the distance between La and Ti atom at the $n$-type interface is 
$\sqrt{3}/2$ times the distance between neighboring Ti 
atoms \cite{Chen-PRB-2009}. Consequently, to the leading
order \footnote{We in fact perform the exact diagonalization and find
that neglecting the hopping between adjacent Ti atoms and other
higher order hopping matrix elements is an excellent
approximation. More importantly, this leading order approximation
described in the text shows that the Ti-La hopping is crucial in
trapping the electrons at the $n$-type interface.}, we neglect the
hopping between adjacent Ti atoms and other higher order hopping
matrix elements. The Hamiltonian becomes block-diagonal and the only
block we need to diagonalize is the leading $2\times2$ Ti-La submatrix
right at the interface:
\begin{equation}
\label{hopping-13} h=\left(\begin{matrix}H_{00} & H_{01} \\
H_{10} & H_{11}\end{matrix}\right)
\end{equation}
Diagonalization of Eq.~(\ref{hopping-13}) with the tabulated values
gives the following eigenvalues and eigenvectors:

\begin{equation}
\label{hopping-14} \varepsilon_1=12.99 \textrm{eV}, \ \ 
|\xi_1\rangle=0.98|\textrm{La}\textrm{-}d_{xy}\rangle-0.18
|\textrm{Ti}^1\textrm{-}d_{xy}\rangle
\end{equation}

\begin{equation}
\label{hopping-15} \varepsilon_2=9.04\textrm{eV}, \ \ 
|\xi_2\rangle=0.18|\textrm{La}\textrm{-}d_{xy}\rangle+0.98
|\textrm{Ti}^1\textrm{-}d_{xy}\rangle
\end{equation}
After diagonalization, we find a pair of bonding and anti-bonding
states at the $n$-type interface. The energy of the binding state
$\varepsilon_2$ is even lower than $H_{22}$, showing that the electrons
prefer the Ti site right at the interface. Therefore, after the
insulating-to-metallic transition occurs, most electrons reside in
this bonding state, which leads to an electron gas bound to the
interface. In other words, the large Ti-La hopping matrix element is
critical in creating a strongly bound state for the electrons at the
interface. Moreover, this hopping is intrinsic to the $n$-type interface 
geometry as it does not significantly change its magnitude before and 
after the polar catastrophe.
This also explains the overall similarity of the electron
distributions at the symmetric double $n$-type superlattice 
(Fig.~\ref{fig:chargespatial}a) 
and the stoichiometric $n$-type
interface (Fig.~\ref{fig:chargespatial}c). Despite differing in the 
lack or presence of oppositely charged carriers and polar fields, the
two systems share the same Ti-La hopping and thus the same
binding force to the interface. 

On the other hand, at the $p$-type interface, holes are found to
occupy bands with essentially pure O-$p$ characters and the maximum of
those bands is located at M
($\textbf{k}=(\frac{\pi}{a},\frac{\pi}{a},0)$). The tight-binding
model is based on the following matrix elements:
\begin{equation}\label{hopping-16} H^{\mu\nu}_{ij}=
\langle \varphi^{\textrm{M}}_{\textrm{O}^i\textrm{-}p_{\mu}}|H|\varphi^{\textrm{M}}_{\textrm{O}^j\textrm{-}p_{\nu}}\rangle
\end{equation}
where $i,j$ refer to different O atoms and $\mu,\nu=x,y,z$ refer to
different O-$p$ orbitals. We calculate all these matrix elements at
the $p$-type interface before and after the polar catastrophe. We find
that the hopping elements between nearest neighbor O atoms are all
close to 0.6 eV, that there is no order of magnitude difference in the
hopping elements, and that there is no discontinuity in going across
the $p$-type interface. This is no surprise because the O atoms form a
continuous network across the $p$-type interface and the hopping
matrix elements do not sensitively depend on the nature of the
surrounding cations. Therefore, the only difference we expect from the
symmetric double $p$-type results (compare Fig.~\ref{fig:chargespatial}b and
Fig.~\ref{fig:chargespatial}d) is that the presence of the polar field at the 
$p$-type interface will drive the holes from the LaAlO$_3$ into the SrTiO$_3$.
However, once in the SrTiO$_3$,
the holes do not feel any strong preference for the interface and therefore 
diffuse further into the SrTiO$_3$ substrate. As our calculations have finite 
film thicknesses,
we expect that the profile of holes resembles the lowest state 
of a free particle in a box and
thus have a maximum density in the middle of the film, as borne out by
Fig. \ref{fig:chargespatial}d.

\begin{table}
\caption{\label{tab:onsite} On-site and hopping
matrix elements of the $n$-type interface before and after the polar
catastrophe, respectively. $l$ is the thickness of LaAlO$_3$. $l=2$ and 
4 unit cells (u.c.) correspond to before and after the polar catastrophe, 
respectively.
We list all the nearest neighbor hopping
matrix elements and the largest next nearest neighbor hopping
matrix element. The first and third columns are the on-site matrix
elements of Ti $d$-states and hopping matrix elements before
the polar catastrophe. The second and fourth columns are the on-site
matrix elements of Ti $d$-states and hopping matrix elements after
the polar catastrophe.}
\begin{center}
\begin{tabular}{c|c|c|c|c|c}
\hline\hline
 \multicolumn{3}{c|}{$H_{ii}$ (eV)} & \multicolumn{3}{c}{$H_{ij}$ (meV)} \\
\hline
$i$     & $l=2$ u.c. & $l=4$ u.c. &  $ij$    & $l=2$ u.c. & $l=4$ u.c. \\
\hline 0  & 12.99  & 12.87  & 0 1   & -697   & -696  \\
\hline 1  & 9.13   & 9.18   & 1 2   & -6.39  & -1.3 \\
\hline 2  & 9.09   & 9.14   & 2 3   & -6.80  & -0.2 \\
\hline 3  & 9.10   & 9.15   & 3 4   & -7.03  & -0.3 \\
\hline 4  & 9.10   & 9.19   & 4 5   & -7.07  & -0.3 \\
\hline 5  & 9.10   & 9.16   & 5 6   & -7.07  & -0.4 \\
\hline 6  & 9.10   & 9.17   & 6 7   & -7.05  & -0.4 \\
\hline 7  & 9.10   & 9.18   & 7 8   & -7.03  & -0.5 \\
\hline 8  & 9.10   & 9.19   & 8 9   & -6.98  & -0.5 \\
\hline 9  & 9.10   & 9.19   & 0 2   & -0.87  & -7.2 \\
\hline\hline
\end{tabular}
\end{center}
\end{table}

\subsection{Oxygen vacancies repulsion from the interfaces}
\label{oxygen-repulsion}

The theoretical results we have discussed to this point have concerned 
ideal interfaces with
sharp boundaries and $1\times1$ in-plane periodicity. An important question
is whether these assumptions are reasonable for describing the
experimentally realized interfaces.  This question is obviously very
difficult to answer in general as the number of possible ways of
perturbing the ideal interfaces is enormous and includes intermixtures,
impurities, vacancies, interstitials, off-stoichiometry, etc. Below,
we focus on answering one simple but important question
for the most prevalent type of imperfections: do oxygen vacancies present
in the SrTiO$_3$ substrate have a strong preference for the SrTiO$_3$
region close to the interface?  Based on both our first
principles results and general physical considerations, the answer
appears to be no. As we show below, the interface SrTiO$_3$ region 
should be relatively free of oxygen vacancies compared to the bulk
SrTiO$_3$ substrate.

\begin{figure}[]
\includegraphics[angle=0,width=12cm]{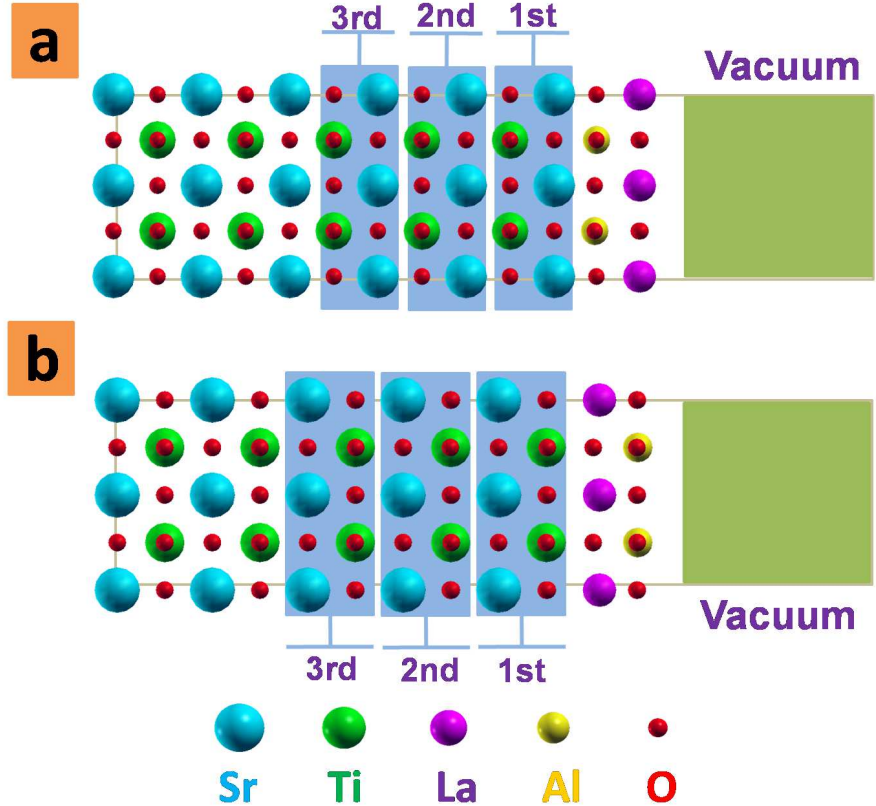}
\caption{\label{fig:ovp2x2}Schematics of the simulation cell for
$\frac{1}{4}$ monolayer oxygen vacancy calculations. One oxygen
vacancy is placed in either SrO and TiO$_2$ of the first, second and
third layer, respectively. \textbf{a)} The $p$-type interface. 
\textbf{b)} The $n$-type interface. }
\end{figure}

Our approach is motivated by experiments that infer a
concentration of one oxygen vacancy per four two-dimensional unit
cells at the $p$-type interface \cite{Hwang-NatMat-2006}.
If one assumes each oxygen vacancy donates fully two electrons, 
this is precisely the amount required to compensate the 
0.5 holes per two-dimensional unit cell from the polar catastrophe. 
The same experiments are also used to infer a
non-zero (but smaller) concentration of oxygen vacancies at the $n$-type
interface \cite{Hwang-NatMat-2006}. To determine the most energetically 
favorable position for oxygen vacancies, we compute 
the formation energy of
one monolayer of oxygen vacancies at the same areal density (1/4 per 
two-dimensional unit cell) at various positions inside the SrTiO$_3$ 
close to the interface
with LaAlO$_3$. We perform calculations on both $p$-type and $n$-type
interfaces using the unit cells shown in Fig.~\ref{fig:ovp2x2}.

Our simulation cells include a $2\times2\times5$ SrTiO$_3$ film and a
$2\times2\times1$ LaAlO$_3$ film and $\simeq$ 25 \AA~of vacuum. The
substrate termination is SrO in order to minimize any
surface effects \cite{Chen-PRB-2009}. 
The SrTiO$_3$ unit cell next to the vacuum is
fixed to simulate the bulk-like substrate and all the other atoms are
relaxed. For a reference, we compute the formation energy of an
isolated oxygen vacancy in bulk SrTiO$_3$ in a
$2\times2\times5$ supercell with one oxygen
vacancy. Starting from the interface, we
place the oxygen vacancy in the SrO or TiO$_2$ layer of the first three
SrTiO$_3$ unit cells and compute the formation energy (see
Fig.~\ref{fig:ovp2x2} for the numbering nomenclature of the layers).
We show the results in Table~\ref{table:ovenergies}. As the oxygen
vacancies move away from the $p$-type interface, the formation energy
decreases, approaching a constant value of 5.4-5.5 eV 
in the bulk-like regions of the SrTiO$_3$ films. The computed bulk 
formation energy is slightly lower (5.18 eV); presumably the small 
difference is due to the fact that the SrTiO$_3$ films in the interface 
geometry are rather thin (only 5 unit cells) and not yet fully in the 
substrate limit. Nevertheless the energetic 
\textit{trend} clearly shows the interface region repels the vacancies. 

Our results show that there is no energetically favorable
binding of oxygen vacancies to either the $p$-type or $n$-type interface. 
A simple physical picture can explain the
repulsion of the oxygen vacancies from both types of interface. An isolated,
neutral oxygen vacancy in SrTiO$_3$ is a donor that binds two electrons
(O has formal charge O$^{2-}$). The electrons reside on localized states 
composed of the $d$-orbitals of the vacancy's surrounding Ti atoms; the
energy of these states is close to the SrTiO$_3$ conduction band edge. 
Thus, oxygen vacancies can be described as a type of hydrogenic system 
with bound electrons. As the vacancy
approaches the interface, the bound electrons experience what are 
essentially hard
wall boundary conditions since the conduction band edge of LaAlO$_3$ is
$\sim$ 2 eV higher than that of SrTiO$_3$.  
As the vacancy approaches the hard wall, its energy increases
due to the electron confinement effect \cite{Satpathy-PRB-1983}. 
Given the basic physical principles behind this argument, it is clear 
that it applies to both
interfaces. Futhermore, it suggests that even though
our first principles results are for relatively high densities of oxygen
vacancies in a particular configuration, 
lower densities and more positionally disordered oxygen vacancies in
the SrTiO$_3$ will still be repelled from the interface region.

The fact that oxygen vacancies are repelled from the interfaces
provides a self-consistent picture for our computations of ideal
interfaces.  Namely, the dominant defects (oxygen vacancies) should
not be present in the immediate vicinity of the interfaces.  For
example, for the $p$-type interface, our theory predicts extended band
states for the holes that diffuse substantially into the SrTiO$_3$
substrate. Since the vacancies are themselves in the bulk of the
substrate, they can trap the holes, rendering them immobile. 
For a relatively thick
SrTiO$_3$ substrate, this situation is likely since only one oxygen
vacancy per four two-dimensional unit cells, distributed over the
three-dimensional volume of the substrate, is required to provide the 
compensating number of electrons.

We would like to comment that experimentally Nakagawa \textit{et al.}
\cite{Hwang-NatMat-2006} found $32 \pm 6\%$ of oxygen vacancies per
two-dimensional unit cell at the $p$-type interface. Their conclusion is
based on a least-square fit of EELS spectra to the reference spectra
of bulk SrTiO$_3$, bulk LaAlO$_3$ and oxygen-deficient SrTiO$_{3-\delta}$ with 
$\delta = 1/4$. The discrepancy between our results and those experiments 
may be
due to the following two non-mutually exclusive reasons. One is that
in our simulations, we include only one unit cell of LaAlO$_3$ owing to
the computational limitations. More realistic simulations would include
thicker LaAlO$_3$ films to simulate the system after the polar catastrophe 
when carriers appear in the vicinity of the interfaces. The presence of
carriers might in principle affect the formation energy of oxygen
vacancies, which needs to be checked in much larger calculations 
in future work. The other reason is the interpretation of
EELS spectra in experiment. The experimental fitting is based on the
assumption that any deviation of O-K edge EELS from the bulk SrTiO$_3$ is
caused by the presence of oxygen vacancies. However, even without
oxygen vacancies at the $p$-type interface, we observe in our calculations 
that the atoms close to the interface move away from their
bulk positions due to the broken symmetry at the interface. The EELS
spectra of the distorted SrTiO$_3$ could be different from that of
the reference bulk structure. Therefore, a more accurate (and more
difficult) fit would take into account the deviation of the EELS spectra
of distorted SrTiO$_3$ from the bulk and other possible factors which
could also affect the EELS spectra, such as intermixture and
nonstoichiometry.

\begin{table}[h!]
\caption{\label{table:ovenergies}
  Formation energies $E$ of $\frac{1}{4}$ monolayer of oxygen
  vacancy at the $p$-type and $n$-type interface. 
  $\frac{1}{4}$ monolayer of oxygen
  vacancy is in either $i$th SrO atomic layer or $i$th TiO$_2$ atomic
  layer.  For reference, the formation energy in the bulk 
  is 5.18 eV.}
\begin{center}
\begin{tabular}{c|c|c|c}
\hline \hline
 \multicolumn{2}{c|}{$p$-type} & \multicolumn{2}{c}{$n$-type} \\
\hline
position of oxygen vacancy & $E$ (eV) & position of oxygen vacancy & $E$ (eV) \\
\hline
1st SrO layer     & 6.25  & 1st TiO$_2$ layer     & 5.65  \\
\hline
1st TiO$_2$ layer & 5.85  & 1st SrO layer         & 5.65  \\
\hline
2nd SrO layer     & 5.69  & 2nd TiO$_2$ layer     & 5.56  \\
\hline
2nd TiO$_2$ layer & 5.56  & 2nd SrO layer         & 5.46  \\
\hline
3rd SrO layer     & 5.60  & 3rd TiO$_2$ layer     & 5.57  \\
\hline
3rd TiO$_2$ layer & 5.50  & 3rd SrO layer         & 5.41  \\
\hline \hline
\end{tabular}
\end{center}
\end{table}

\subsection{Thickness dependence of sheet carrier density}
\label{DFT-thickness}

In experiment, the sheet carrier density
was observed to display a thickness dependence, 
but a range of results exist. The sheet carrier density
depends on both sample growth conditions and post-annealing 
\cite{Chen-Adv-2010}. 
In this section, we present first-principles calculations of
the sheet carrier density for different LaAlO$_3$ film thicknesses. 
We study the $np$-type, $n$-type and QW systems, and discuss the similarities
between these different interface systems.

\begin{figure}[t]
\includegraphics[angle=0,width=10cm]{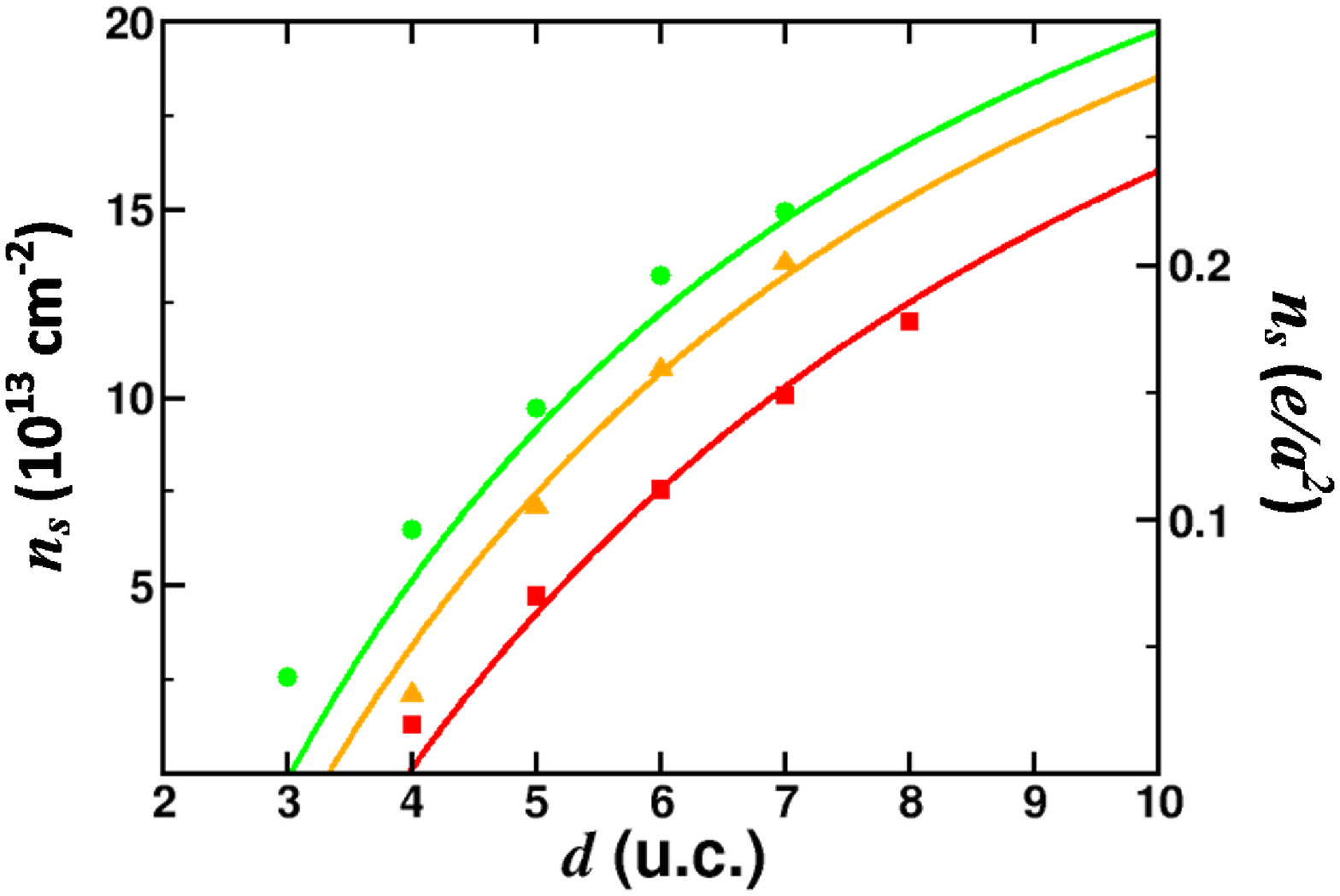}
\caption{\label{fig:charge} Comparison of the sheet carrier 
density computed with the continuous model (see Section \ref{model-thickness}) 
and DFT simulations. The red, orange and green lines
are the model calculations of the QW, the $n$-type and 
the $np$-type interfaces, respectively. The
squares, triangles and circles are the DFT results of the QW, 
the $n$-type and the $np$-type interfaces, respectively.}
\end{figure}

\begin{figure}[t]
\includegraphics[angle=0,width=10cm]{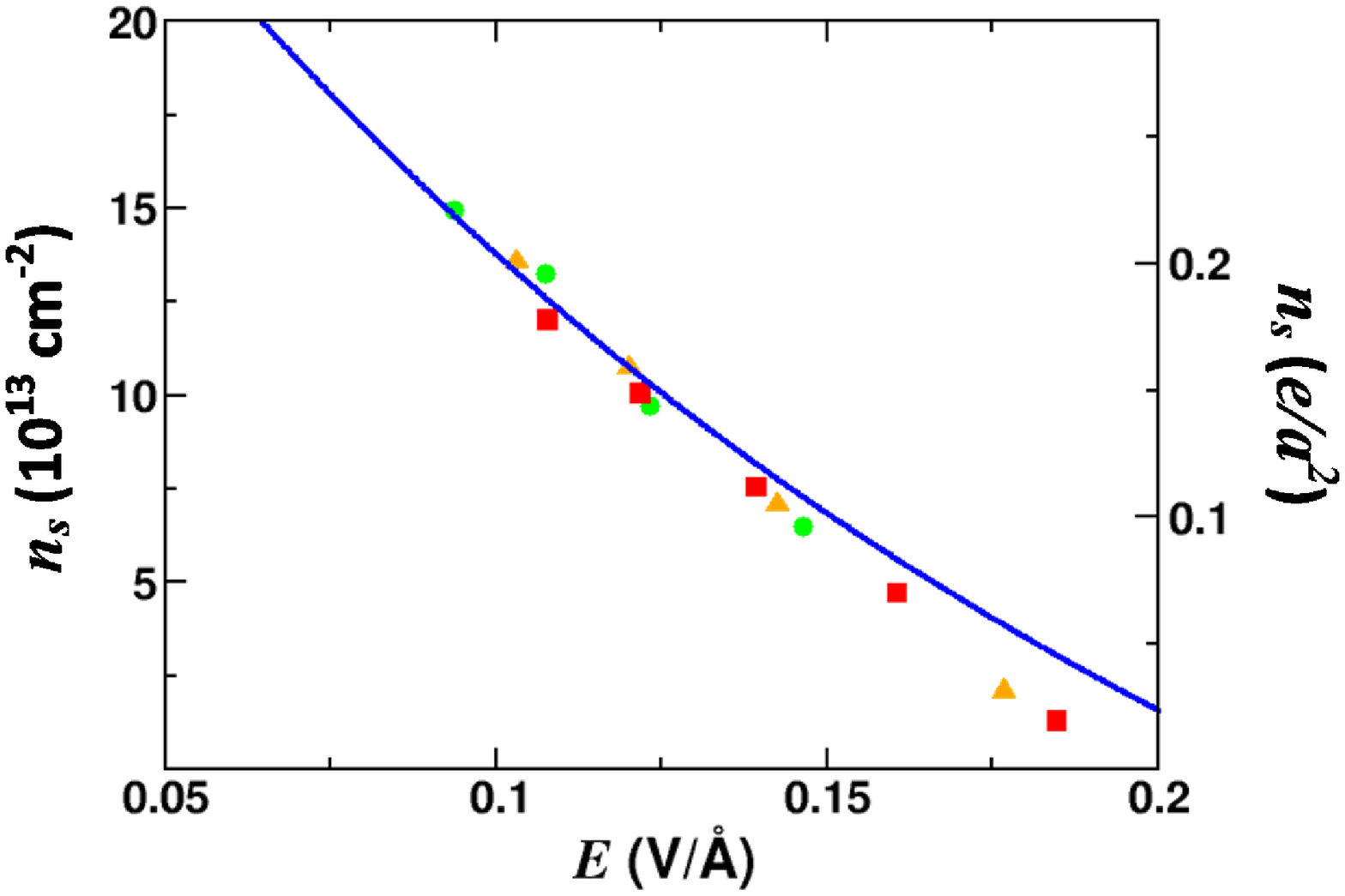}
\caption{\label{fig:nsvsE} The sheet carrier density versus the
internal electric field through the LaAlO$_3$ thin film. The red
squares, orange triangles, and green circles are for the quantum well,
the $n$-type, and the $np$-type interfaces, rspectively. The blue line is
the result of the continuous model described in Section \ref{model-thickness} .}
\end{figure}

We define the `sheet carrier density' as the sum of all the conduction
electrons per square unit cell in the Ti $d$-states of the SrTiO$_3$
substrate (for $n$-type and $np$-type interfaces) or in the Ti
$d$-states of the single embedded TiO$_2$ layer (for the QW
systems). Fig. \ref{fig:charge} shows the sheet carrier density at
different thicknesses for all three types of interface. The results of
$n$-type interface are similar to what Son {\it et al.}  recently
obtained \cite{Son-PRB-2009}. Below the critical separation in each
system, there are no mobile carriers since the system is
insulating. Above the critical separation, the sheet carrier densities
increase in all three cases, following a similar manner. In order to
further reveal the similarities among these interface systems, we show
in Fig. \ref{fig:nsvsE} the relation of the sheet carrier density
$\sigma_s$ versus the internal electric field $E$ through the
LaAlO$_3$.  The internal electric field is determined as follows.  We
calculate the macro-averaged smoothed total potential along the (001)
direction. The internal electric field is 
the slope of the smoothed potential.  Though the geometry details
differ, the figure shows that the $(E, \sigma_s)$ relations are almost
identical for all three interface types. This suggests that a single
continuous model will be able to describe the behavior of all three
interfaces. We discuss such a model in Section \ref{model-thickness}.

We note that in our calculations of the sheet carrier density,
we include all of the electrons in the conduction bands, 
regardless of whether they in fact contribute to the observed
conductivity or not.  We find that after the critical separation, the sheet
carrier density increases continuously with the thickness of the
LaAlO$_3$ film, asymptoting to 0.5 $e/a^2$.  This behavior
qualitatively agrees with some experiments \cite{Huijben-NatMat-2006,
Sing-PRL-2009}, but conflicts with others
\cite{Mannhart-Science-2006}, in which the sheet carrier density 
remains almost constant (15 times smaller than 0.5 $e/a^2$) after the
critical separation. It is possible that the low sheet carrier
density in Ref. \cite{Mannhart-Science-2006} may be explained in terms
of Anderson localization induced by disorder \cite{Satpathy-PRL-2008, 
Son-PRB-2009}.
 
\subsection{External Field Effect}
\label{DFT-external}

\begin{figure}[]
\includegraphics[angle=-90,width=14cm]{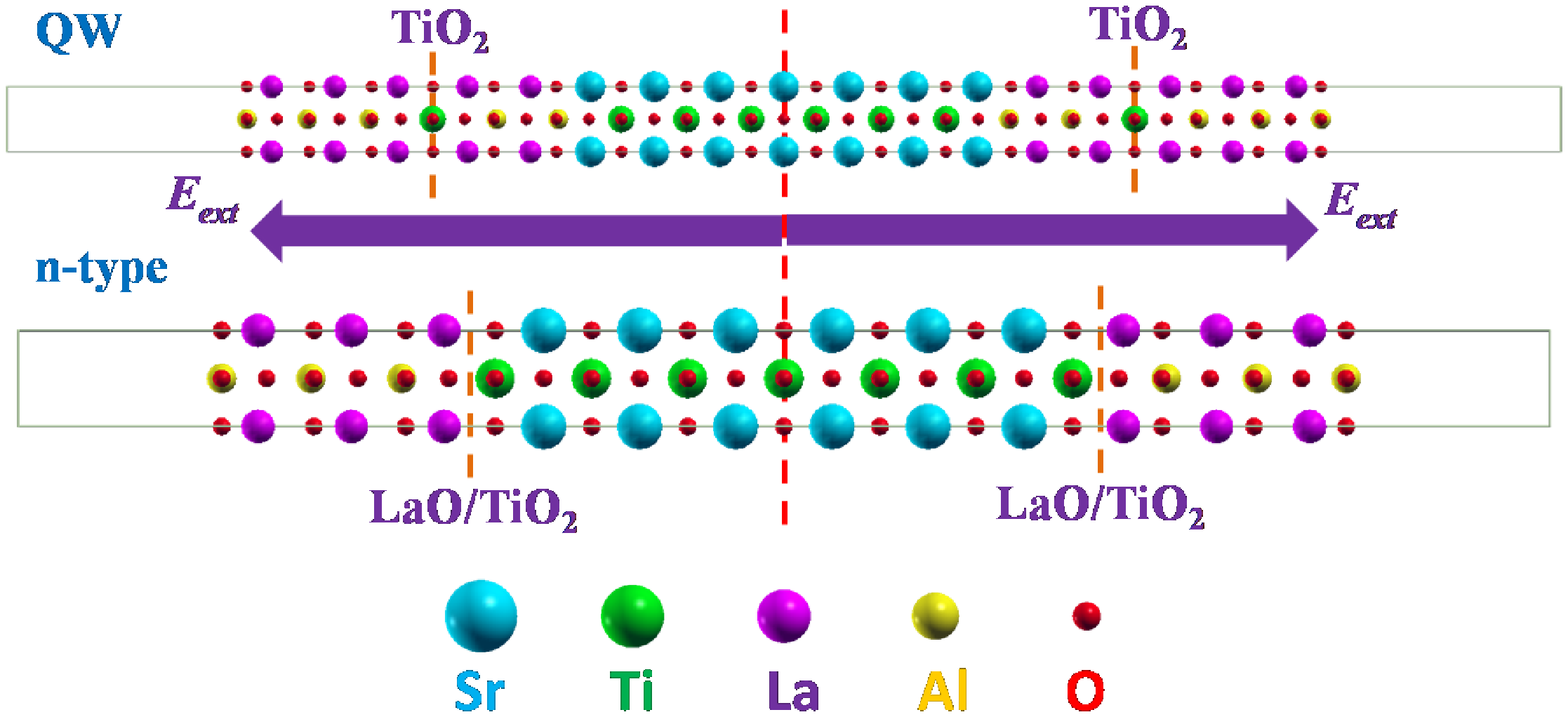}
\caption{\label{fig:fieldunitcell}
The supercells used for the field effect
calculations. For the QW and the $n$-type interface, both simulation cells
are mirror-symmetric. The external electric field is also
mirror-symmetric. The dashed line in the middle is the
symmetry axis. The embedded TiO$_2$ layer in the QW and the $n$-type
interface LaO/TiO$_2$ are also highlighted by the dashed line.}
\end{figure}

We have shown that there is an insulating-to-metallic transition as a
function of \lao thickness both in the SrTiO$_3$/LaAlO$_3$ interface
and the quantum well (QW) systems. In this section, we examine the effects of
inducing this transition at subcritical thicknesses via an applied
electric field.  We focus on the $n$-type interface and the QW system
with 3 u.c. of LaAlO$_3$ (for both systems, the 
critical separation is 4 u.c. in the DFT simulations). 
To study the field effect, we
apply a homogeneous electric field perpendicular to the
interface. In order to avoid artificial effects from periodic
boundary conditions (see Appendix \ref{appendix-PBC}), we apply
two electric fields of equal magnitude and opposite direction to each
half of a symmetric simulation cell (see Fig. \ref{fig:fieldunitcell}).
Since the system is mirror symmetric, for simplicity we only focus
on half of the simulation cell. At the $n$-type interface, the external
electric field is parallel to the intrinsic electric field
through the LaAlO$_3$ film, and in the QW systems, the direction 
is parallel to the field through the LaAlO$_3$ capping layer. 
Based on the polar catastrophe picture, such external electric fields 
will add to intrinsic electric fields and further reduce the 
energy gap, so that the insulating-to-metallic 
transition will occur at a subcritical separation.

We performed the simulations with different magnitudes of external
electric field on both the $n$-type interface and QW systems.  We need to 
elucidate a subtle point here: in the QW, the external
electric field reduces the energy difference between the Ti $d$-states
of the embedded TiO$_2$ layer and O $p$-states on the surface. This is
the `Ti-O energy difference' introduced in Section
\ref{sec:quantumwell}. Once this energy difference disappears, the
insulating-to-metallic transition occurs. This energy difference is
well defined because we can identify two Bloch states, one 
which characterizes the Ti $d$-states of the embedded TiO$_2$ layer
and the other which characterizes the O $p$-states on the surface, 
and simply take their energy difference.  At the $n$-type interface, the
external electric field also reduces the energy difference between
the Ti $d$-states of the first Ti atom at the interface and the O
$p$-states on the surface. However, this energy difference is not 
easily identified because the external electric field also
exists throughout the SrTiO$_3$ substrate, so the Bloch states 
that accept the transferred electrons are now composed of a mixture 
of all the Ti $d$-states in the SrTiO$_3$ thin film. For the purposes 
of this section, we simply extract the energy gap of this $n$-type 
interface system (between the O $p$-states on the surface and the lowest 
occupied Ti $d$-states) versus the external field. In Section 
\ref{model-external}, we will discuss more fully the relation of this 
energy gap and what would happen in an actual experiment. 

\begin{figure}[]
\includegraphics[angle=-90,width=9cm]{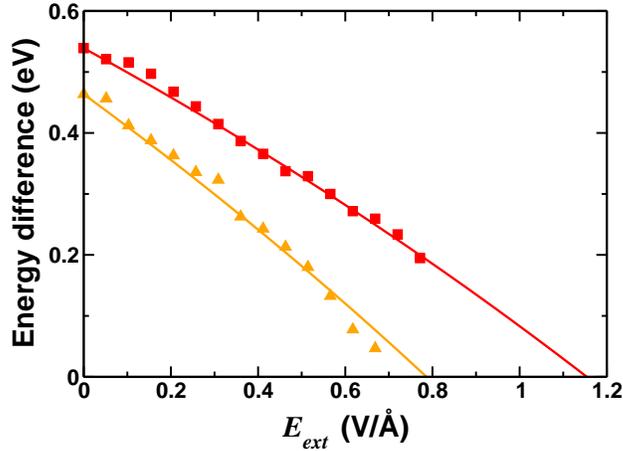}
\caption{\label{fig:externalfield}
Energy differences versus the external
electric field.  The red squares and orange triangles correspond to the 
DFT-computed Ti-O energy difference for the QW systems and 
the energy gap of the $n$-type interface, respectively.
The solid lines are the results of the continuous model described in
Section \ref{model-external}.}
\end{figure}

In Fig. \ref{fig:externalfield} we show the computed energy gap of 
the $n$-type interface and the Ti-O energy difference of the QW as 
a function of external electric field. As the figure shows, both 
monotonically decrease with increasing external electric field. Thus, the DFT 
simulations demonstrate how an external electric field can 
induce an insulating-to-metal transition. 

\section{Continuous model}
\label{model}

In order to shed more light on the nature of the LaAlO$_3$/SrTiO$_3$
interface, we develop a simple model which approximates the LaAlO$_3$
film as a homogeneous continuous medium. 
This model shows that the polar
and dielectric properties of LaAlO$_3$ largely determine the thickness
dependence of the sheet carrier density and 
the external field effect behavior.

\subsection{Thickness dependence of sheet carrier density}
\label{model-thickness}

In this section, we use model calculations, aided by DFT results,
to quantitatively describe the insulating-to-metallic transition
that occurs at the $n$-type and $np$-type interfaces and
the quantum well (QW) systems. As detailed below, we find that the electric
field dependence of $\epsilon_L$ is critical in obtaining quantitative 
accuracy.

\begin{figure}[]
\includegraphics[angle=0,width=9cm]{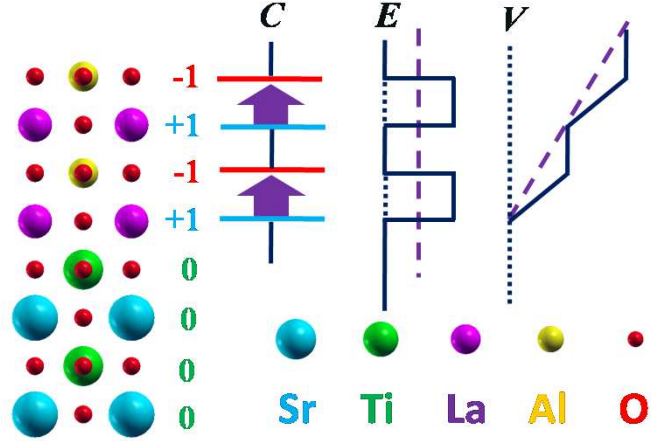}
\caption{\label{fig:average} Schematic illustration of the continuous model.
In the ionic limit, the LaAlO$_3$ thin film can be considered as a
serial connection of capacitors. The electric field through
LaAlO$_3$ is like an impulse and the resulting potential takes a
stair shape. The dashed purple lines are the average electric field and
the average potential.}
\end{figure}

Before the insulating-to-metallic transition occurs, the alternating
positively charged (LaO)$^+$ and negatively charged (AlO$_2$)$^-$
layers in the LaAlO$_3$ film can be idealized as a series of
capacitors (see Fig. \ref{fig:average}).  As the figure
illustrates, the thickness of each capacitor is only half of the
unit cell of LaAlO$_3$ (there is zero voltage drop across the
other half). Therefore the inner electric field through the LaAlO$_3$
behaves like an impulse and the resulting potential takes the shape
of a stair function. If we approximate the LaAlO$_3$ thin film as a
continuous homogeneous media and average this electric
field, we can determine the average internal electric field:

\begin{equation}
\label{charge-1}\overline{E}=\frac{1}{2}\frac{4\pi
\sigma_0}{\epsilon_L}=\frac{4\pi \overline{\sigma}}{\epsilon_L},
\end{equation}
where $\sigma_0=e/ a^2$ is an electron per two-dimensional unit
cell, $\epsilon_L$ is the dielectric constant of LaAlO$_3$, and

\begin{equation}
\label{charge-2}\overline{\sigma}=\frac{\sigma_0}{2}=\frac{e}{2a^2}.
\end{equation}
Once the insulating-to-metallic transition occurs, some electrons
transfer across the interface to counteract the internal electric
field through the LaAlO$_3$ thin film. The new average internal
electric field through the LaAlO$_3$ film is then given by:

\begin{equation}
\label{charge-3}\overline{E}=\frac{4\pi(\overline{\sigma}-\sigma_s)}{\epsilon_L},
\end{equation}
where $\sigma_s$ is the transferred electron density, or sheet carrier density.
It is easy to recast Eq.~(\ref{charge-3}) into:

\begin{equation}
\label{charge-4} \sigma_s(\overline{E})=\overline{\sigma}-
\frac{\overline{E}\epsilon_L}{4\pi},
\end{equation}
where $\overline{\sigma}=0.5e/a^2$. We also take into account that
the dielectric constant $\epsilon_L$ depends on the internal
electric field $\overline{E}$. The field dependence can be phenomenologically
described by the Landau theory, and takes the following approximate form
\cite{Devo-Phil-1949, Antons-PRB-2005} (also see Appendix \ref{appendix-field}):

\begin{equation}
\label{charge-5}
\epsilon_L(\overline{E}) \simeq \epsilon_0
\left(1+\left(\frac{\overline{E}}{\mathcal{E}_0}\right)^2\right)^{-1/3},
\end{equation}
where $\epsilon_0$ is the dielectric constant of LaAlO$_3$ at
vanishing electric field and $\mathcal{E}_0$ is a characteristic electric
field. Eq.~(\ref{charge-4}) and Eq.~(\ref{charge-5}) establish a
relation between the internal electric field $\overline{E}$ through the 
LaAlO$_3$ and the sheet carrier density $\sigma_s$, both of which 
can be calculated independently from DFT simulations. 
We also performed a separate slab calculation (see Appendix 
\ref{appendix-field-lao} for details) 
and determined the parameters $\epsilon_0$ and $\mathcal{E}_0$ 
in Eq. (\ref{charge-5}) as:
\begin{equation}
\label{charge-6}\epsilon_0=40.95, \phantom{5} \mathcal{E}_0=0.15
\textrm{V/\AA}
\end{equation}
Fig. \ref{fig:nsvsE} compares $(E, \sigma_s)$ computed via
DFT and the model. 
We can see from Fig. \ref{fig:nsvsE} that $(E, \sigma_s)$ 
almost lies on the same curve in all three cases, which 
does not depend on the details
of the structure. The continuous model shows a good
agreement with the DFT results, and as the internal electric field
decreases (i.e. the thickness of the LaAlO$_3$ increases), the
continuous model becomes almost exact. This is expected because in
the limit of an infinitely thick LaAlO$_3$ film, the interface and
surface regions become neglegible.
Our results show that even with only 7 u.c. of LaAlO$_3$ 
({\it i.e.} the internal electric field $<$ 0.12 V/\AA), the
continuous model already works very well.

However, it is not easy to directly measure the internal electric
field through the LaAlO$_3$ experimentally because the LaAlO$_3$ film
is only a few unit cells thick. It is more useful to relate the
sheet carrier density to the nominal thickness of the LaAlO$_3$ film
({\it i.e.} the number of LaAlO$_3$ unit cells grown on top of the SrTiO$_3$
substrate). Therefore, we need to find how the internal electric
field through the LaAlO$_3$ depends on the LaAlO$_3$ thickness.

\begin{figure}[]
\includegraphics[angle=-90,width=9cm]{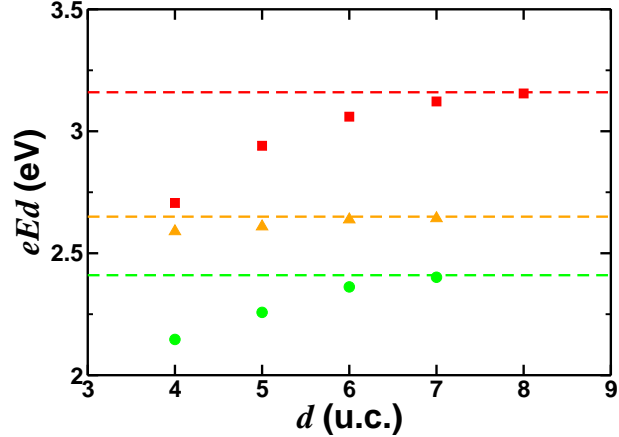}
\caption{\label{fig:K} The potential difference across the LaAlO$_3$
thin film, defined by $e\overline{E}d$, of the quantum well (QW), 
the $n$-type and the $np$-type interfaces. Squares, triangles, 
and circles correspond to the QW, the $n$-type, and the $np$-type 
interfaces, respectively.
$\overline{E}$ is the average internal electric field
along the (001) direction through the LaAlO$_3$ thin film and $d$ is the
nominal number of LaAlO$_3$ unit cells (note that the
SrTiO$_3$-strained LaAlO$_3$ lattice constant
$a_{\rm LAO}=0.953a_{\rm STO}$). The dashed lines highlight the data point 
of the largest thickness in each interface configuration.}
\end{figure}

When the polar catastrophe takes place at the $n$-type
interface, electrons leave the surface O-$p$ states and are transfered
into the Ti-$d$ states. The charge transfer is halted once a common
Fermi level is reached. Approximately, this means that the Ti-$d$
conduction band edge of the SrTiO$_3$ substrate is at the same energy
as the O-$p$ surface valence band edge. Looking back at
Fig. \ref{fig:interface}, we can see that there is a remanent field
through the LaAlO$_3$ film and thus a potential difference. The
existence of the electric field is also visible in
Fig. \ref{fig:nsvsE} from our calculations. 
The potential difference across the LaAlO$_3$ thin film, denoted by $K$, 
can be related to the remanent internal electric field by:
\begin{equation}
\label{charge-7}e\overline{E}d= K
\end{equation}
where $d$ is the thickness of LaAlO$_3$.
In principle $K$ could depend on the thickness of LaAlO$_3$. But
the detailed calculations show (see Fig. \ref{fig:K}) 
that $K$ quickly saturates to a constant as the thickness $d$ gets larger.
This could be understood because $K$
is essentially determined by the energies of the relevant electronic
states and the band offsets, which are interface properties and have 
little dependence on the thickness of LaAlO$_3$.
We therefore approximate $K$ as a
constant. We illustrate later that this approximation does
not significantly change the physics in the continuous model.

Inserting Eq.~(\ref{charge-7}) into Eq.~(\ref{charge-4}) and 
Eq.~(\ref{charge-5}), we eliminate $\overline{E}$ and obtain:

\begin{equation}
\label{charge-8}\sigma_s=\overline{\sigma}-\frac{K}{4\pi
ed}\epsilon_0\left(1+\left(\frac{K}{e\mathcal{E}_0d}\right)^2\right)^{-1/3}
\end{equation}
Eq.~(\ref{charge-8}) highlights some qualitative features of the sheet carrier
density. The critical separation is the
smallest $d$ which makes $\sigma_s > 0$. Below the
critical separation, the sheet carrier density is zero. Above the
critical separation, the sheet carrier density gradually increases
and then saturates to $\overline{\sigma}=0.5e/a^2$  in the limit of large
$d$.

The parameter $K$ in Eq.~(\ref{charge-8}) is determined as follows.  We
simulate $n$-type, $np$-type interfaces and QW with different
thicknesses and calculate the macro-averaged smoothed electric field through
the LaAlO$_3$ at each thickness. In Fig. \ref{fig:K}, we show the
potential difference across the LaAlO$_3$ thin film, defined by
$e\overline{E}d$.   
We can
see that in all three cases, $e\overline{E}d$ quickly saturates as $d$
gets larger. Since we do not have a microscopic model to describe how
the potential difference $e\overline{E}d$ changes with the thickness
$d$, we use the last data (corresponding to the largest $d$) to
determine the values of $K$, which are highlighted by the dashed line in
Fig. \ref{fig:K}.  In each case, we find that

\begin{equation}
\label{charge-10} K_{\rm QW}=3.16 \textrm{ eV}, \phantom{5} K_{n}=2.64
\textrm{ eV}, \phantom{5} K_{np}=2.40 \textrm{ eV}.
\end{equation}
The $K$ in all three cases turn out to be different from each other 
and larger than the DFT-computed band gap of SrTiO$_3$ (1.85 eV).
This is partly because the band offsets of these three cases are not 
the same. However, the small valence band offsets alone 
\cite{Albina-PRB-2007, Pickett-PRB-2008} can not explain the big difference 
between the $K$ value and SrTiO$_3$ band gap. A more important factor 
is that we treat $\overline{E}$ as a uniform electric field throughout 
the LaAlO$_3$ thin film. However, close to the interface and surface, 
the internal electric field should be different from that 
in the middle of the LaAlO$_3$ film but difficult to average 
in those regions. Therefore we use an approximate relation ($K=e\overline{E}d$) 
to determine the potential difference, but as shown later, this approximation 
is good enough to reproduce the DFT-calculated thickness dependence of 
sheet carrier density. 
         
Equipped with $K$, $\epsilon_0$ and $\mathcal{E}_0$, we can calculate the sheet
carrier density at different thicknesses using Eq.~(\ref{charge-8}).
We compare the sheet carrier densities calculated from the continuous
model with those from the DFT simulations. The results are shown in
Fig. \ref{fig:charge}.
The agreement is good in all three cases even though
Eq.~(\ref{charge-7}) is not exact.

We would like to comment that recently Son \textit{et al.} 
\cite{Son-PRB-2009} also calculated the thickness dependence 
of sheet carrier density and they made a similar model by assuming 
that the dielectric constant of LaAlO$_3$ has no field dependence 
and is simply constant. Rather 
than Eq.(\ref{charge-8}), they had a simpler formula

\begin{equation}
\label{charge-11}\sigma_s=\overline{\sigma}\left(1-\frac{d_c}{d}\right)
\end{equation}
where $\overline{\sigma}$ and $d_c$ are two constants determined by 
fitting. They found $\overline{\sigma}=0.455 e/a^2$, close to the ideal 
value $0.5 e/a^2$. Our analysis is different in that we fix 
$\overline{\sigma}=0.5 e/a^2$, which is necessitated by the polar catastrophe 
mechanism, but take into account that the dielectric constant of LaAlO$_3$ 
has a strong field dependence, as confirmed by the DFT calculations.

\subsection{External electric field}
\label{model-external}

In order to get a more quantitative description of the external field effect,
we apply the continuous model established in the previous
section. In principle, at the subcritical separation (3 u.c.),
the continuous model should break down because the LaAlO$_3$ film
is so thin that
the interface and surface regions cannot be neglected and
the LaAlO$_3$ is no longer uniform. However, we simplify the situation
by assuming that LaAlO$_3$ is still a homogeneous media with
the dielectric constant given by Eq.~(\ref{charge-5}) with the parameters
Eq.~(\ref{charge-6}), but has an effective thickness $d_{eff}$. 
In order to simplify our notations, we use $E$ instead of 
$\overline{E}$ to denote the averaged macroscropic fields.

Based on the above argument, for the quantum well systems we have:

\begin{equation}
\label{field-1} \Delta=\Delta_0-(E^{L}-E^{L}_0)d^{L}_{eff},
\end{equation}
where `$L$' stands for LaAlO$_3$. $\Delta$ is the Ti-O
energy difference at a given $E_{ext}$ and $E^{L}$
is the internal electric field. $\Delta_0$ and $E^{L}_0$ indicate 
the values of $\Delta$ and $E^{L}$, respectively, at vanishing 
external electric field:

\begin{equation}
\label{field-2} E^L=\frac{4\pi
\overline{\sigma}+E_{ext}}{\epsilon_L(E^L)}
\end{equation}
and 

\begin{equation}
\label{field-3} E^L_0=\frac{4\pi
\overline{\sigma}}{\epsilon_L(E^L_0)}
\end{equation}
Equations~(\ref{field-1}, \ref{field-2}, and \ref{field-3}) establish
the relation between $\Delta$ and $E_{ext}$. The effective thickness
cannot be given $a$ $priori$ from the model. Instead, we fit the data
and find $d^{L}_{eff}=5.9 \textrm{\AA} = 1.6$ u.c. A comparison of
$(E_{ext}, \Delta)$ for the quantum well systems
  using DFT calculations and the model calculations
 is shown in Fig. \ref{fig:externalfield}. 

For the $n$-type interface, we also need to account for the potential drop 
in the SrTiO$_3$ due to the external electric field. Therefore we have:

\begin{equation}
\label{field-4}
\Delta=\Delta_0-(E^{L}-E^{L}_0)d^{L}_{eff}-E^{S}d^{S}_{eff}
\end{equation}
where `$S$' stands for SrTiO$_3$. $\Delta$ is the energy gap of the
system and $\Delta_0$ is the value of $\Delta$ at vanishing external electric
field. Note that in the absence of an external electric field, the
SrTiO$_3$ film is unpolarized in the continuous model. $E^{S}$ and
$E_{ext}$ are related by:

\begin{equation}
\label{field-5} E^S=\frac{E_{ext}}{\epsilon_S(E^S)}
\end{equation}
We use the Berry phase method \cite{Souza-PRL-2002} 
to calculate $\epsilon_S (E)$ and
fit the data using Eq.~(\ref{charge-5}) (see Appendix 
\ref{appendix-field-sto} for details). The results of the fit are:

\begin{equation}
\label{field-6}
\epsilon^{S}_0=309.6, \mathcal{E}^{S}_0=49.2 \textrm{V/}\mu\textrm{m}
=4.92 \times 10^{-3} \textrm{V/\AA} 
\end{equation}
We find that the values of $d^{L}_{eff}$ in different systems are different. 
Hence, in principle we have two fitting parameters in Eq.~(\ref{field-4}):
$d^{L}_{eff}$ and $d^S_{eff}$. However, it turns out that
Eq.~(\ref{field-4}) very insensitively depends on $d^{S}_{eff}$.  This
is expected because the dielectric constant of SrTiO$_3$ is at least 10
times larger than that of LaAlO$_3$ and the nominal thicknesses of
LaAlO$_3$ and SrTiO$_3$ are both 3 u.c. in the DFT calculations (see
Fig. \ref{fig:fieldunitcell}), and thus the potential drop across the
SrTiO$_3$ is much smaller than that across the LaAlO$_3$ film.
Therefore we just use the real thickness of SrTiO$_3$  $d^{S}_{eff}=11.1$ 
\AA~(from the relaxed structure) and fit the data with $d^L_{eff}$. We find:
$d^{L}_{eff}=7.5$ \AA $\simeq 2$ u.c. Fig. \ref{fig:externalfield} compares 
the values of $(E_{ext}, \Delta)$ of the $n$-type interface computed by 
DFT and the model, respectively.

Fig. \ref{fig:externalfield} shows that the 
critical electric fields $E^c_{ext}$ extracted from the DFT calculations 
for the $n$-type interface and the QW are 0.79 and
1.15 V/\AA, respectively. In the continuous model, the critical voltage 
across the whole sample is given by:

\begin{equation}
\label{external-field}V_{c}=
\frac{E^{c}_{ext}+4\pi\overline{\sigma}}{\epsilon_{\textrm{LAO}}}
d_{\textrm{LAO}}+\frac{E^{c}_{ext}}{\epsilon_{\textrm{STO}}}d_{\textrm{STO}}
\end{equation}
where $d_{\textrm{LAO}}$ and $d_{\textrm{STO}}$ are 
the thickness of LaAlO$_3$ and SrTiO$_3$, respectively. 
$\overline{\sigma}=0.5 e/a^2$ and $E^{c}_{ext}$ is the critical 
external electric field. In order to estimate the critical 
voltage for the experimental setup, we need to use the 
thickness of the SrTiO$_3$ substrate in experiment. 

\begin{figure}[]
\includegraphics[angle=0,width=14cm]{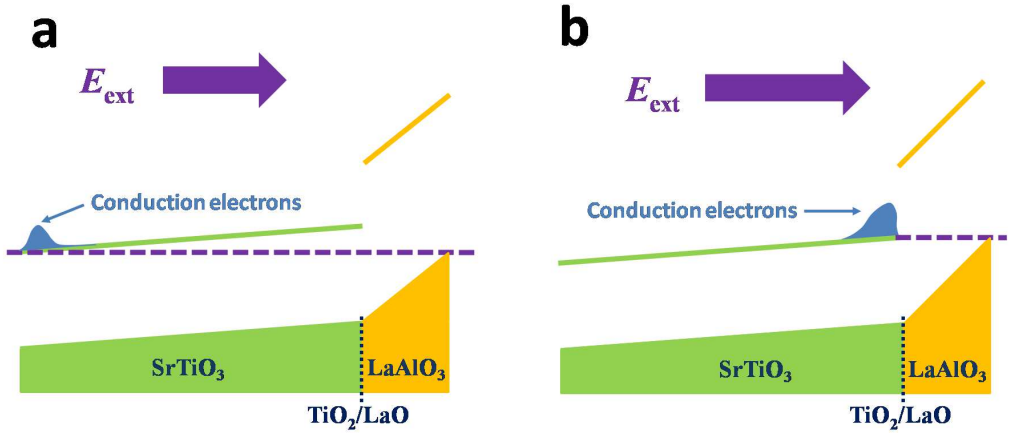}
\caption{\label{fig:external-field} \textbf{a)} The spatial 
distribution of conduction electrons in the DFT simulations.
The conduction electrons occupy the lowest energy states available, 
which are located at the bottom surface. 
\textbf{b)} The spatial distribution of conduction electrons 
in the actual experiment. The conduction electrons get trapped 
at the $n$-type interface by the tunneling barrier due to 
the self-consistent potential well formed at the interface 
that largely stems from the Ti-La hopping.
The length of arrows illustrates the magnitude of external field.}
\end{figure}

Here we need to clarify some subtleties: in the presence 
of applied external field, DFT simulations do not 
realistically represent the spatial distribution of 
conduction electrons. Fig. \ref{fig:external-field} illustrates 
the reason for the discrepancy. 
In the DFT simulations, the conduction electrons fill the 
lowest energy states. Since the ``external'' field bends 
the SrTiO$_3$ conduction bands, these states are located 
at the bottom surface of the SrTiO$_3$ substrate, as shown
schematically in Fig. \ref{fig:external-field}a. Although 
in principle the experimental situation is the same, realistically, 
electrons can only tunnel a few unit cells from the surface; thus they
get trapped at the $n$-type interface, as illustrated in 
Fig. \ref{fig:external-field}b. The trapping mechanism is the 
Ti-La interfacial hopping, which lowers the energy of the Ti 
atom closest to the interface and creates the energy barrier through 
which the electrons must tunnel. 
Hence, the real critical field has to be 
larger than the computed one, so that the energy of surface states 
equals that of the
interface states (Fig. \ref{fig:external-field}b). The critical 
external field $E_{ext}^c$ we obtain from DFT simulations 
is therefore only a lower bound.    

Now we make a simple estimation of the lower bound: 
at low temperatures ($< 5$ K), the dielectric constant of SrTiO$_3$ can 
be as large as $2.5\times 10^4$ \cite{Saifi-PRB-1970} 
and the thickness of the SrTiO$_3$ 
substrate is typically $\simeq 1$ mm. Considering that 
$\epsilon_{\textrm{LAO}} \simeq 30$ and $d_{\textrm{LAO}}\simeq 10$~\AA, 
the second term in Eq.(\ref{external-field}) is dominant. Therefore, 
for the $n$-type interface, we have:

\begin{equation}
\label{external-field-ntype}
V^{n\textrm{-type}}_{c}\simeq\frac{E^{c}_{ext}}{\epsilon_{\textrm{STO}}}
d_{\textrm{STO}}\simeq 320~\textrm{V}
\end{equation}
and for the QW, 

\begin{equation}
\label{external-field-QW}
V^{\textrm{QW}}_{c}
\simeq\frac{E^{c}_{ext}}{\epsilon_{\textrm{STO}}}d_{\textrm{STO}}\simeq 480 
~\textrm{V}
\end{equation}
As these estimates are based on data
from DFT simulations in which the band gap of SrTiO$_3$ is
underestimated, they provide only lower
bounds on the critical voltage. Experimentally, the critical 
voltage of the $n$-type interface
with a 1 mm thick SrTiO$_3$ substrate has been measured to be 
$\simeq$ 70 V at low temperature \cite{Mannhart-Science-2006}, which is
much smaller than our estimates. The discrepancy may imply that the
simple model of an ideally sharp interface is not sufficient to model 
the external field doping, due to additional details that 
play an important role at the $n$-type interface. Various types 
of atomic reconstructions, such as cation intermixtures at the interface 
\cite{Hwang-NatMat-2006} or
oxygen vacancies on the surface \cite{Cen-NatMat-2008} may possibly trigger 
an insulating-to-metallic transition, leading to a smaller critical 
voltage. 

\section{Conclusion}
\label{conclusion}

We present a detailed study of ideal LaAlO$_3$/SrTiO$_3$
interfaces and a new class of quantum wells. In both
systems, we confirm an intrinsic insulating-to-metallic transition. The
observed transition can be triggered either by thickening the
LaAlO$_3$ layers or by imposing an external electric field, both of
which can be explained by the polar catastrophe mechanism. 
We show that for both the $n$-type interface and QW, the
realistic critical separation (taking into account the underestimation
of DFT band gap) is around 6 unit cells. 
We also show that, given a typical \sto substrate 
thickness of 1mm, the lower bound for the critical voltage necessary to 
induce an insulating-to-metallic transition is 
estimated to be $\sim 300$ V for the ideal $n$-type interface 
and $\sim 500$ V for the QW. In addition to theoretically demonstrating 
the observed physical properties, we provide a microscropic explanation of 
the observed binding of conduction electrons at the $n$-type
interface, a phenomenon which can not be described by the polar 
catastrophe mechanism alone. We demonstrate that
the large La-Ti hopping matrix
element at the $n$-type interface, 
which is absent in both bulk
constituents and is unique to the $n$-type interface,
lowers the energy of the Ti atom at the interface relative to all other 
Ti sites, thus binding the electrons.
Futhermore, we develop a continuous model that captures the
essence of polar structure and dielectric properties and predicts the
thickness dependence of sheet carrier densities. 

We conclude by discussing some of the outstanding issues related to this 
system. Our understanding of the atomically sharp interfaces and
QW suggests that the electronic properties of
LaAlO$_3$/SrTiO$_3$ interfaces observed in experiments are unlikely to be 
solely a result of electronic reconstructions based on the polar catastrophe
mechanism. For example, the theoretical critical separation, predicted by
the polar catastrophe picture, is 6 
u.c. \cite{Chen-PRB-2009, Demkov-PRB-2008, Pickett-PRL-2009,
  Son-PRB-2009}, which is larger than the experimental value (4 u.c.) 
\cite{Mannhart-Science-2006}. In addition, while the 
calculated thickness dependence of the sheet carrier density agrees 
qualitatively with some experiments \cite{Huijben-NatMat-2006,Sing-PRL-2009}
in that the sheet carrier density increases with LaAlO$_3$ thickness, it
disagrees with other experiments \cite{Mannhart-Science-2006} which show
an almost constant sheet carrier density above the critical
separation. The sheet carrier density we calculate here includes all
the conduction electrons. Whether some of them may be prone to the
Anderson localization \cite{Satpathy-PRL-2008} is an issue beyond the scope 
of the DFT calculations. However, the possibility of multiple type of carriers
\cite{Seo-2009} is an interesting question that warrants further
study in both theory and experiment. Another remaining puzzle is that 
the polar catastrophe
predicts the coexistence of electrons at the interface and holes one
the surface, while experiments find that the surface region is 
insulating and only electron-like carriers are detected in the
transport measurement \cite{Mannhart-Science-2006}. Recently, x-ray 
photoemssion measurement found that the polar field through LaAlO$_3$ is much 
smaller than the theoretical predictions \cite{Segal-PRB-2009}. 
It is possible that
various atomic reconstructions, such as cation intermixting,
nonstoichiometry and defects could also significantly affect the
electronic structures of these interface systems and be partially
responsible for the discrepancies mentioned above. While a detailed
discussion of atomic reconstructions is beyond our present work, it 
serves as an interesting direction for future research.  

\appendix

\section{Conduction electron and hole densities}
\label{appendix-density}

In this appendix, we explain how we calculate the \textit{conduction} 
electron and hole densities for the symmetric superlattices and stoichiometric 
interface systems. The basic tool we use is the local density of states (LDOS)
for the system 

\begin{equation}
\label{localdos}D(\textbf{r},E)=\sum_{n\textbf{k}}
|\psi_{n\textbf{k}}(\textbf{r})|^2\delta(E-\epsilon_{n\textbf{k}})
\end{equation}
which we integrate over the appropriate energy range. 

\subsection{Double $n$-type and $p$-type superlattices}

The symmetric nonstoichiometric double $n$-type ($p$-type) superlattice 
is shown in Fig. \ref{fig:symif}a (Fig. \ref{fig:symif}b). The corresponding 
electron (hole) denisty is presented in Fig. \ref{fig:chargespatial}a 
(Fig. \ref{fig:chargespatial}b).  
For the symmetric superlattices, there is no polar field and analysis
of the density of states shows that there is an easily
identifiable energy gap at all locations in the film; the Fermi level
is either above the conduction band edge for the the $n$-type superlattice
or below the conduction band for the $p$-type superlattice.
For the transferred electron density (Fig. \ref{fig:chargespatial}a), 
we integrate the LDOS from 
the middle of the band gap to the Fermi level, 

\begin{equation}
\label{doublee}\eta(\textbf{r})=\int^{E_F}_{\cal{E}}D(\textbf{r},E)dE
\end{equation}
where $\cal{E}$ is the energy in the middle of the band 
gap. We are in fact counting all the electrons in the conduction bands. 
For the transferred 
hole density (Fig. \ref{fig:chargespatial}b), similarly we integrate the 
LDOS from the Fermi level to the middle of the band gap, 

\begin{equation}
\label{doublee}\xi(\textbf{r})=\int^{\cal{E}}_{E_F}D(\textbf{r},E)dE
\end{equation}
\textit{i.e.} counting all the holes in the valence bands. 
The integrated LDOS $\eta(\textbf{r})$ and $\xi(\textbf{r})$ are 
then averaged over the $xy$ plane and finally plotted along the $z$ 
direction.

\subsection{Stoichiometric $n$-type and $p$-type interface systems}

The stoichiometric $n$-type ($p$-type) interface system is shown in 
Fig. \ref{fig:interface}b (Fig. \ref{fig:interface}c). The corresponding  
\textit{conduction} electron (hole) density is shown in Fig. 
\ref{fig:chargespatial}c (Fig. \ref{fig:chargespatial}d).  
For the stoichiometric $n$-type and $p$-type interfaces, the polar field
ensures that the bands edges in the LaAlO$_3$ are not flat and that the
Fermi level will intersect the band edges of the SrTiO$_3$ as well as the
surface of the LaAlO$_3$ film (see Fig. \ref{fig:interface}). 
However, the local density of states in the
SrTiO$_3$ film still clearly shows a band gap. Therefore, we can also
compute the transferred charges at the interface 
using the above formula. The only
complication is that in addition to showing the transferred charges in
the interface region, $\eta(\textbf{r})$ and $\xi(\textbf{r})$ 
will necessarily have
contributions localized at the surface region of LaAlO$_3$ which are not of
direct interest when studying the interface alone; the choice of axis
range in Fig. \ref{fig:chargespatial}c and \ref{fig:chargespatial}d 
effectively excludes this contribution. As
a consistency check, we can also compute the transferred charge by
computing the atomic projections of all Bloch states in the system and
identifying all those states with the proper atomic character: Ti-$d$ or La-$d$
character and partial occupancy for the $n$-type interface, and O-$p$
character and partial occupancy for the $p$-type interface.  One can then
manually sum up these particular contributions to get the electron and
hole distributions, and the results agree with the previous method in 
the relevant regions.

\section{Periodic boundary condition effects}
\label{appendix-PBC}

When we perform a slab calculation, the materials in the simulation
are generally polarized either due to intrinsic polar fields ({\it
  e.g.} LaAlO$_3$) or external electric fields. In DFT calculations,
the periodic boundary conditions (PBC) imposed on the simulation cell
influence the screening properties of the materials. We will show in the
following two sections that the presence of PBCs does not significantly 
affect the polarization due to intrinsic polar fields in the system, 
but it does induce large
artificial effects when one studies the response of the system to an external
field, requiring a careful set up of the simulation and interpretation 
of the results.

\subsection{Polarization due to intrinsic fields}
\label{appendix-PBC-intrinsic}

In this section, we illustrate that in the SrTiO$_3$/LaAlO$_3$ interface
system, PBCs induce an artificial field through SrTiO$_3$, but we 
show that this effect is neglegible.   

As shown in Fig. \ref{fig:PBCeffect}a, the two boundary conditions of 
displacement $\mathbf{D}$ at each interface give:

\begin{equation}
\label{app-1}
D_{\rm LAO}-D_{\rm STO}=4\pi\sigma
\end{equation}

\begin{equation}
\label{app-2}
D_{\rm STO}-D_{\rm V}=0
\end{equation}
PBCs require that the total potential drop over the whole simulation cell 
is equal to zero:

\begin{equation}
\label{app-3}
E_{\rm STO}s+E_{\rm LAO}l+E_{\rm V}v=0
\end{equation}
where $s$, $l$ and $v$ are the thicknesses of SrTiO$_3$, LaAlO$_3$ and vacuum, 
respectively. Considering that $E_i=D_i/\epsilon_i$ 
($i$ is  SrTiO$_3$, LaAlO$_3$ or vacuum),
we can solve $D_{\rm STO}$ and $D_{\rm LAO}$ explicitly,

\begin{equation}
\label{app-4}
D_{\rm STO}=-4\pi\sigma\frac{l/\epsilon_{\rm LAO}}
{s/\epsilon_{\rm STO}+l/\epsilon_{\rm LAO}+v/\epsilon_{\rm V}}
\end{equation}

\begin{equation}
\label{app-5}
D_{\rm LAO}=4\pi\sigma\frac{s/\epsilon_{\rm STO}+v}
{s/\epsilon_{\rm STO}+l/\epsilon_{\rm LAO}+v/\epsilon_{\rm V}}
\end{equation}
The dielectric constants of SrTiO$_3$, LaAlO$_3$ and vacuum are roughly:

\begin{equation}
\label{app-6}
\epsilon_{\rm STO}\simeq 300 \gg \epsilon_{\rm LAO} \simeq 25 \gg 
\epsilon_{\rm V}=1
\end{equation}
and the thicknesses of SrTiO$_3$, LaAlO$_3$ and vacuum in the simulation cell 
are on the same order of magnitude (within a factor of two), therefore we 
have the following simplifications:

\begin{equation}
\label{app-7}
E_{\rm LAO}\simeq \frac{4\pi\sigma}{\epsilon_{\rm LAO}}
\end{equation}

\begin{equation}
\label{app-8}
E_{\rm STO}\simeq -\frac{4\pi\sigma}
{\epsilon_{\rm LAO}\epsilon_{\rm STO}}\frac{l}{v}=-\frac{l}
{v\epsilon_{\rm STO}}E_{\rm LAO}\to |E_{\rm STO}|\ll |E_{\rm LAO}|
\end{equation}

\begin{equation}
\label{app-9}
E_{\rm V}\simeq -\frac{4\pi\sigma}
{\epsilon_{\rm LAO}}\frac{l}{v}=-\frac{l}
{v}E_{\rm LAO}\to E_{\rm V}v \simeq -E_{\rm LAO}l
\end{equation}
Eq.~(\ref{app-8}) and Eq.~(\ref{app-9}) show that though 
SrTiO$_3$ is artificially polarized due to the PBCs, the 
electric field through SrTiO$_3$ is much smaller than 
the intrinsic field through LaAlO$_3$ and the voltage 
built across the LaAlO$_3$ film mostly drops in the vacuum. 
Based on the above approximate dielectric constants, the field 
in SrTiO$_3$ is only $\sim 0.3\%$ of that in LaAlO$_3$. In addition, 
we can further reduce the field in SrTiO$_3$ by increasing the thickness 
of the vacuum $v$; the above relations give a quantitative measure of the 
(small) error for any finite value of $v$.

\subsection{Polarization due to an applied external electric field}
\label{appendix-PBC-extrinsic}

We illustrate in this section that when we use a slab calculation 
and a sawtooth potential to simulate an
external electric field through nonpolar materials ({\it e.g.} SrTiO$_3$), 
PBCs will artificially undermine the screening and give rise to a 
significantly smaller dielectric constant. 

Fig. \ref{fig:PBCeffect}b shows a
schematic of how an external electric field $E_{ext}$ along the $z$
direction is screened in a nonpolar material. There are two
induced electric fields. The one in the material is the stardard
depolarization field $E_{dep}$ and the other field, which we call 
the outside field $E_{out}$, is in the
vacuum with the opposite direction. 
$E_{ext}$, $E_{dep}$ and $E_{out}$ are the $magnitudes$
of each field; their directions are explicitly shown in Fig.
\ref{fig:PBCeffect}b. We denote the size of the unit cell along 
the $z$ direction by
$L$ and the thickness of the slab by $d$. The sawtooth potential
automatically satisfies the periodic boundary condition (the
reversed part is not shown in Fig. \ref{fig:PBCeffect}b). 
Therefore, the induced electric fields required to satisfy the 
periodic boundary condition are given by:

\begin{figure}[]
\includegraphics[angle=-90,width=9.5cm]{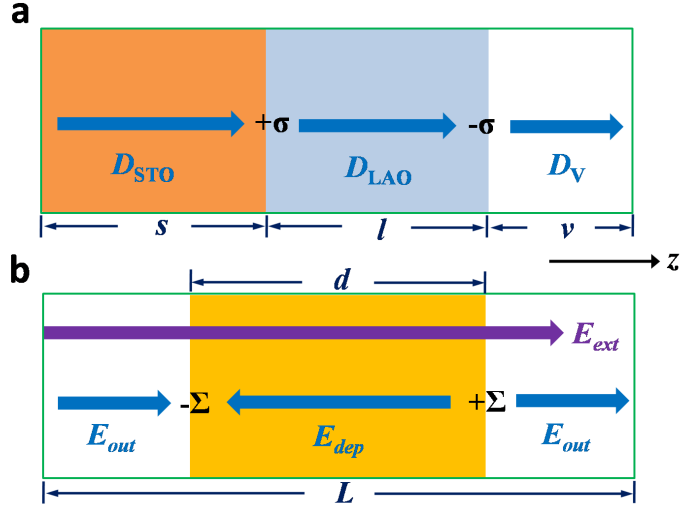}
\caption{\label{fig:PBCeffect}Schematics of how the periodic boundary
condition affects the screening of the electric fields.
\textbf{a)} Polarization due to intrinsic fields. The shaded 
parts are SrTiO$_3$ and LaAlO$_3$, respectively. The interface is 
$n$-type. The unshaded part denotes vacuum. \textbf{b)} Polarization 
due to an applied electric field. The shaded part is a general nonpolar 
material and the empty part is vacuum.}
\end{figure}

\begin{equation}
\label{app-10}
E_{dep}d=E_{out}(L-d)
\end{equation}
On the other hand, Gauss's law gives:

\begin{equation}
\label{app-11}
4\pi \Sigma = E_{dep}+E_{out}
\end{equation}
where the surface charge density $\Sigma$ is related to the
polarization in the material by

\begin{equation}
\label{app-12}
\Sigma = \textbf{n}\cdot\textbf{P}=P
\end{equation}
and the polarization $P$ is related to the total electric field $E_{tot}$ by

\begin{equation}
\label{app-13}
P=\chi E_{tot}
\end{equation}
where $\chi$ is the permitivity. From Fig. \ref{fig:PBCeffect}b, 
it is easy to see that

\begin{equation}
\label{app-14}
E_{tot}=E_{ext}-E_{dep}
\end{equation}
Combining Eq.~(\ref{app-10}-\ref{app-14}) gives

\begin{equation}
\label{app-15}
E_{tot}=\frac{E_{ext}}{1+4\pi\chi\left(1-\frac{d}{L}\right)}
\end{equation}
Now we can identify the dielectric constant $\epsilon$

\begin{equation}
\label{app-16}
\epsilon=\frac{E_{ext}}{E_{tot}}=1+4\pi\chi\left(1-\frac{d}{L}\right)
\end{equation}
Eq.~(\ref{app-16}) is different from the familiar formula 
$\epsilon=1+4\pi\chi$ in
that we have an extra factor $(1-d/L)$. In typical slab calculations
$(1-d/L) \sim 30\%$ and for high-$k$ materials ({\it e.g.} SrTiO$_3$) 
$\chi\gg 1$ so that $\epsilon$ is dominated by the second term 
in Eq.~(\ref{app-16}) and thus the reduction factor is a significant error.

\subsection{A remedy}
The origin of the deviation of the dielectric constant from 
the correct value of $1+4\pi\chi$
is that the electric field induced by the bound charge $\pm \Sigma$ does not
completely serve as the depolarization field, but is instead split into two
parts, $E_{dep}$ and $E_{out}$. The presence of $E_{out}$ is purely due to the
imposed periodic boundary condition, and is unphysical. However, 
in practical slab calculations it is too computationally expensive 
to make $d/L$ small.

In order to get rid of $E_{out}$, we need a simulation cell in which the
depolarization field automatically satisfies periodic boundary
conditions. Therefore, we use a mirror-symmetric simulation cell in which the
external electric field is also mirror-symmetrically distributed. By
symmetry, $E_{dep}$ automatically satisfies periodic boundary
conditions and $E_{out}$ is guaranteed to be zero in the vacuum. Thus,
the external electric field is correctly screened, but with a relatively 
modest increase in computational expense due to doubling the simulation 
cell along one direction.

\section{Field dependence of dielectric constant}
\label{appendix-field}

The Landau theory phenomenologically describes the field dependence
of the dielectric constant \cite{Devo-Phil-1949, Antons-PRB-2005}. 
It is assumed that the free energy of the
system $F(P,T)$ can be expanded in even powers of the polarization $P$:

\begin{equation}
\label{app-17}
F(P,T)=F_0+AP^2+BP^4+CP^6+...
\end{equation}
where the coefficients $A,B,C...$ may depend on the temperature $T$.
Keeping only terms in $F$ to the fourth order, we obtain:

\begin{equation}
\label{app-18}
E=\frac{\partial F}{\partial P}=2A P + 4BP^3
\end{equation}
and the permitivity follows:

\begin{equation}
\label{app-19}
\frac{1}{\chi}=\frac{\partial E}{\partial P}=2A+12BP^2
\end{equation}
Eq.~(\ref{app-18}) and Eq.~(\ref{app-19}) uniquely determine $\chi=\chi(P)$
and $P=P(E)$. The analytical solution to Eq.~(\ref{app-18}) $P=P(E)$
is complicated, but we can find a useful interpolation scheme 
\cite{Devo-Phil-1949, Antons-PRB-2005} by noting that as $P$ is small, 
$\chi \to (2A)^{-1}$ 
and as $P$ is large, $\chi \propto P^{-2}$ and $E \propto P^3$, 
that is $\chi \propto E^{-2/3}$, so

\begin{equation}
\label{app-20}
\chi\simeq\chi_0\left(1+\left(\frac{E}{\mathcal{E}_0}\right)^2\right)^{-1/3}
\end{equation}
For high-$k$ materials, we can also approximate 
$\epsilon \simeq 4\pi \chi \gg 1$, giving the final expression:

\begin{equation}
\label{app-21} 
\epsilon\simeq\epsilon_0\left(1+\left(\frac{E}{\mathcal{E}_0}\right)^2\right)^{-1/3}
\end{equation}
where $\epsilon_0$ and $\mathcal{E}_0$ are fitting parameters.

We note that the truncation of Eq.~(\ref{app-17}) is
based on the assumption that $P$ is small (\textit{i.e.} $E$ is small). However,
the Landau theory itself does not give a characteristic polarization
or electric field. Instead, we consider $\mathcal{E}_0$ as a
characteristic electric field whose value is determined by the
fitting. Thus, data points much larger than $\mathcal{E}_0$ should
not be used in the fitting because Eq.~(\ref{app-17}) would break
down. This is a self-consistent check.

\subsection{LaAlO$_3$}
\label{appendix-field-lao}

\begin{figure}[t]
\includegraphics[angle=-90,width=9.5cm]{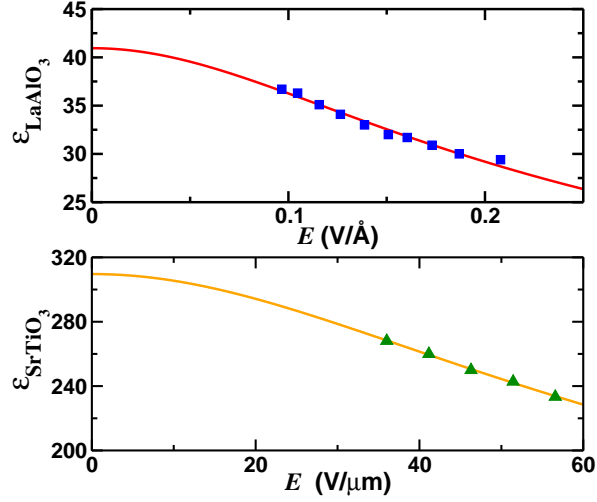}
\caption{\label{fig:dielectric}Electric field dependence of dielectric
constants of LaAlO$_3$ and SrTiO$_3$. The solid squares are the results of
the slab calculations and the solid triangles are the results of Berry phase
calculations. The solid lines are the fitting results using
Eq.~(\ref{app-21}).}
\end{figure}

Before we calculate the dielectric constant $\epsilon_L$ of LaAlO$_3$, 
we need to elucidate a subtle point. The
$\epsilon_L$ we calculate here is only well defined in the thin
slab of LaAlO$_3$, not in the bulk. This is because we are in the
region of very strong electric field ($\sim 0.7$~V/$a_{\rm LAO}$). With a few
unit cells, the potential difference built across the slab will 
be larger than the band gap of LaAlO$_3$, and Zener tunneling will
occur. Bulk LaAlO$_3$ can not accommodate such a
large electric field and the Berry phase method \cite{Souza-PRL-2002}
for calculating the
dielectric constant breaks down in this regime. Instead, we resort to a
mirror-symmetric slab calculation (see Fig. \ref{fig:LAO-slab} and 
Appendix \ref{appendix-PBC}). 
We turn on a mirror-symmetric external electric field $E_{ext}$ and calculate
the macro-averaged electric field $E_{tot}$ through the
material. The dielectric constant is then defined as:

\begin{figure}[]
\includegraphics[angle=-90,width=14cm]{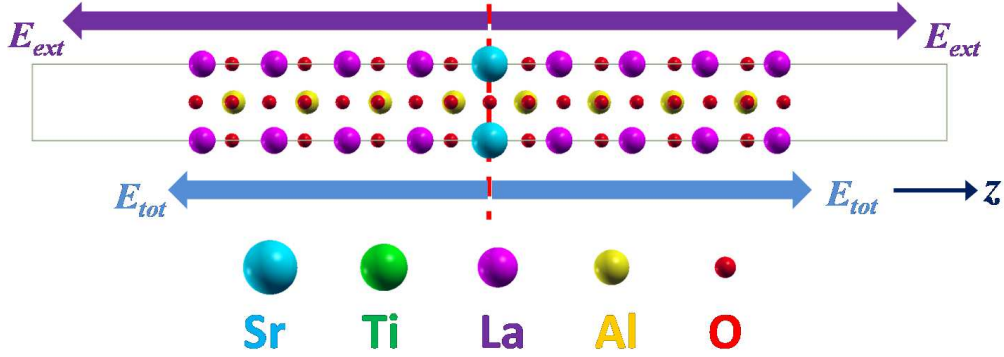}
\caption{\label{fig:LAO-slab}Schematics of the simulation cell of LaAlO$_3$
slab calculations. Both the simulation cell and the
external electric field are mirror-symmetric. All fields through
LaAlO$_3$ are along the $z$ direction (perpendicular to the 
interface).}
\end{figure}
 
\begin{equation}
\label{app-22}
\epsilon_L(E_{tot})=\frac{E_{ext}}{E_{tot}}
\end{equation}
The raw data and the fitting curve using Eq.~(\ref{app-21}) are
shown in Fig. \ref{fig:dielectric}. The fitting results are:

\begin{equation}
\label{app-lao}\epsilon_0=40.95, \phantom{5} \mathcal{E}_0=0.15
\textrm{V/\AA}
\end{equation}

\subsection{SrTiO$_3$}
\label{appendix-field-sto}

Since SrTiO$_3$ is a nonpolar material with a large dielectric
constant, the typical internal electric field through SrTiO$_3$ is
much smaller than that through LaAlO$_3$. The dielectric constant we
are interested in can be defined in the bulk (based on the
argument of metastable states) and calculated using the Berry
phase method. In order to accurately determine the atom positions,
we use $6\times6\times20$ $k$-point sampling and lower the force
convergence threshold to 8 meV/ \AA. We directly calculate the
total polarization $P_{tot}$ (both ionic and electronic) in the
unit cell at a given total electric field $E_{tot}$ and the
dielectric constant follows straightforwardly:

\begin{equation}
\label{app-23}
\epsilon_S(E_{tot})=4\pi\frac{P_{tot}}{\Omega E_{tot}}+1
\end{equation}
where $\Omega$ is the volume of unit cell. The raw data and
fitting curve are shown in Fig. \ref{fig:dielectric}, with 
the fitting results:

\begin{equation}
\label{app-sto}\epsilon^{S}_0=309.6, \mathcal{E}^{S}_0=49.2 
\textrm{V/}\mu\textrm{m}=4.92 \times 10^{-3} \textrm{V/\AA} 
\end{equation}

\bibliography{sto-lao-long-prb11}

\end{document}